\documentstyle[aps,prl,eqsecnum,preprint,tighten]{revtex}
\begin{document}
\title
{Quantum phase transitions in
frustrated  two-dimensional antiferromagnets}
\author{Andrey V. Chubukov${}^{1,2,3}$, Subir Sachdev${}^{2}$ and  T.
Senthil${}^{2}$}
\address{
${}^{1}$Department of Physics, University of Wisconsin, Madison,
WI 53706\\
${}^{2}$Departments of Physics and Applied Physics, P.O. Box 208120,\\
Yale University, New Haven, CT 06520-8120\\
and ${}^{3}$P.L. Kapitza Institute for
Physical Problems, Moscow, Russia}
\date{January 5, 1994}
\maketitle
\begin{abstract}
We study frustrated, two-dimensional, quantum antiferromagnets
in the vicinity of a quantum transition from a non-collinear,
magnetically-ordered ground state to a quantum disordered phase.
The general scaling properties of this transition are described.
A detailed study of a particular field-theoretic model of the
transition, with bosonic spin-1/2 spinon
fields, is presented.
Explicit universal scaling forms for a variety of
observables are obtained and the results are compared with numerical data on
the spin-1/2 triangular antiferromagnet.
Universal properties of an alternative field-theory, with confined spinons,
are also briefly noted.
\end{abstract}
\narrowtext

\section{introduction}

There has been a remarkable recent revival of interest in the
low-energy properties of two-dimensional ($2d$)
frustrated quantum antiferromagnets. In
part, this interest was triggered by the discovery of strong magnetic
fluctuations in the
high-$T_c$ superconductors; however, frustrated magnetic systems
are interesting in their own right,
in the light of numerous
theoretical predictions on the nature of disordered
ground states in quantum
spin systems~\cite{Kalm_Laugh,Sach-Read,Girvin,Ar-Auerb}.

Three kinds of frustrated $2d$ systems have been
studied intensively, both experimentally and theoretically.
First, are antiferromagnets on a triangular lattice such as
$VCl_{2}, VBr_{2}, C_{6}Eu, NaTiO_2$ etc~\cite{Harris}.
Theoretical studies of such
antiferromagnets go back to 1973 when Anderson and Fazekas~\cite{And-Fasek}
first suggested that for $S=1/2$, quantum fluctuations may be strong enough to
destroy the classical $120^0$ ordering of Heisenberg spins. Though
most of the subsequent numerical and analytical studies do
indicate~\cite{trianglro} the presence of long-range order at zero
temperature ($T$),
these studies also show~\cite{Singh-Huse}
 that quantum fluctuations are quite strong.

A second frustrated system is the antiferromagnet on a kagome lattice. It is
believed
to describe the second layer of $^{3}He$ on graphite~\cite{Elser} and
$SrCr_{8-x}Ga_{4+x}O_{19}$ and related compounds~\cite{expkag}. The
effects of quantum fluctuations in kagome
antiferromagnets are far stronger than in triangular ones~\cite{Kag},
 and numerical studies of $S=1/2$ systems support a
quantum-disordered ground state at $T=0$~\cite{kagt=0,Singh-Huse}.
Besides, large $S$ kagome antiferromagnets  display the
Villain order from disorder phenomenon
{}~\cite{Villain}: in the semiclassical approximation,
they possess a strong `accidental' degeneracy
which is lifted only by the zero-point motion of quantum
 spins~\cite{Kag,swkag}.
Tunneling between a sequence of nearly degenerate ground states
(which differ in energy only due to quantum  fluctuations), may
also contribute substantially  to the reduction in the strength of the large
$S$ long-range order~\cite{Chris}.

Finally, there are also studies of antiferromagnets on the square lattice
which are frustrated by adding second and third neighbor
couplings~\cite{Sach-Read,J1J2}. These systems
show interesting phases with incommensurate, planar, spiral correlations.

A key feature of the local spin correlations in the systems above,
which will be crucial in our analysis,
is that they are {\em non-collinear\/}.
Unlike the unfrustrated square lattice, the spins
are not locally either parallel or anti-parallel to one another.
The analysis in this paper will mostly assume that the spins are
{\em coplanar\/} although this second restriction is mostly in the interests
of simplicity.

So far  we have discussed the situation at $T=0$. Experiments, however, are
performed at finite $T$ when thermal fluctuations are also present. The
effects of thermal fluctuations for $2d$ Heisenberg systems are well
known~\cite{2DfiniteT} - they
destroy long-range magnetic correlations at arbitrary small $T$.
Suppose, first, that  the ground state is nearly perfectly ordered.
It is clear, then, that at small $T$,
 thermal fluctuations will be significantly
 more important than quantum fluctuations, and the low-$T$ behavior
will be predominantly classical - the primary
effect of quantum fluctuations will
be a renormalization of the couplings at $T=0$.  This is the
low-$T$ ``renormalized-classical'' regime
which was studied in detail in Ref~\cite{CHN}, and later
observed~\cite{expren_cl} in a number of experiments on  undoped
square-lattice antiferromagnets at sufficiently
low $T$. Consider, next, the physics when
the system is quantum-disordered at $T=0$. Then all
thermally induced fluctuations are suppressed by a (presumed) spin-gap at low
enough temperatures and
and the low-energy dynamics is purely quantum mechanical
- this is the
``quantum-disordered''~\cite{CHN} regime.

However, there is a third, intriguing possibility which
arises when the ground state of the system is not too far from
a $T=0$, second-order quantum transition between the magnetically-ordered
and quantum disordered states.
Then it is easily possible to find the so-called ``quantum-critical''
regime where classical and thermal fluctuations are equally important.
This is a \underline{high} temperature regime with respect to any
energy-scale which measures the deviation of the ground state of the
antiferromagnet from the quantum transition point; on the magnetically-ordered
side a convenient choice for this energy-scale
 is a spin stiffness, $\rho_s$.
However it is also a \underline{low} temperature regime with respect to
a microscopic, short-distance energy scale like a nearest-neighbor
exchange constant, $J$.
If the couplings are precisely critical, then the
quantum-critical region
stretches down to lowest $T$
- this is unlikely to be realized in antiferromagnets without fine-tuning of
an external parameter {\em e.g.\/} pressure or doping.
 However, even if the system
is not precisely at the critical point, but $T$ is larger than
$\rho_s$ on the ordered side,
 or a corresponding energy scale $\Delta$
on the disordered side, we still observe
essentially quantum-critical behavior because at such $T$ we effectively probe
the system at scales where it
 does not know on which side of the transition it will end up in its ground
state.
However,  if the long-range order at $T=0$ is very well established
(or, if on the quantum-disordered side, $\Delta$ is very large)
the condition $k_B T > \rho_s$ ($k_B T > \Delta$)
for quantum-criticality may conflict
or interfere with $k_B T < J$ and the
the quantum-critical behavior can be
overshadowed by nonuniversal short-range fluctuations.
Thus we require that $\rho_s$ ($\Delta$) be reasonably small, and then
then we may expect to observe quantum-critical behavior at $T$ smaller than
$J$.

In a recent publication with J. Ye~\cite{CSY}, two of us
considered whether a
quantum-critical
region exists in the square-lattice $S=1/2$
antiferromagnet. We computed various experimentally measurable quantities
such as the uniform susceptibility, the correlation length,
the dynamic structure factor,
and the spin-lattice relaxation rate for antiferromagnets with collinear spin
correlations. We
found reasonable agreement between the
quantum-critical results and the experimental data~\cite{expren_cl}
 on $La_{2-x}Sr_{x}CuO_4$ and with
numerical results  on $S=1/2$ antiferromagnets~\cite{numsl}.
We argued, therefore, that this system is  quantum-critical at intermediate
temperatures.
In frustrated $2d$ systems, quantum fluctuations are
likely to be
far stronger. It is therefore
 reasonable to expect that quantum-critical behavior
may be observed in frustrated systems as well. In the present paper, we
will present detailed predictions about the quantum-critical properties
of frustrated antiferromagnets with non-collinear correlations,
to help elucidate this possibility.

The study of quantum-critical behavior is not the only purpose of our
analysis. We will also consider the behavior of various observables in the
renormalized-classical region. Previous studies in this this region were
performed by Azaria {\em et. al.\/}~\cite{Azaria}, who
focused on a renormalization group analysis for the correlation length.
Below, we present, for the
first time,
expressions for the uniform susceptibility and dynamic structure factor
of renormalized-classical, non-collinear antiferromagnets.

An important issue, which makes a study of non-collinear antiferromagnets
considerably
more difficult than collinear ones, is that the nature of the
quantum-disordered
phase and the universality of the transition are not well established.
The large-$N$ $Sp(N)$ theories~\cite{Sach-Read,trieste} have argued that
the quantum-disordered phase of non-collinear antiferromagnets has deconfined,
spin-1/2, bosonic spinons.  In this paper, we derive a macroscopic
field-theoretical model which has the same behavior, and
study the universality class
of the transition between a quantum-disordered phase with deconfined spinons
and the magnetically
ordered state. We find, quite generally, that such a transition is in the
universality class of the $O(4)$-vector model in spacetime dimension $D=3$.
 This result agrees with the semiclassical renormalization group analysis
of the magnetically ordered side
in $D=2+\epsilon$
dimensions of Azaria {\em et. al\/}~\cite{Azaria}.
 We will then go on to determine numerous
universal, finite temperature properties of such antiferromagnets. These
properties have many striking differences from those of the collinear
antiferromagnets~\cite{CSY} which possessed confined spinons.
Note however, there are other treatments of the transition in non-collinear
antiferromagnets~\cite{Kawamura} which do not have $O(4)$ exponents at $D=3$.
 We will review
these in Appendix~\ref{confined}
and show that they in fact have confined spinons.
The universal magnetic properties of these approaches differ only in a minor
way from those of Ref~\cite{CSY} and will therefore not be discussed in any
detail.

We will begin in Section~\ref{ordparm} by defining carefully, and with
considerable
generality, the order parameter of coplanar
antiferromagnets~\cite{AM,Dombre_Read,Azaria}. We will also express the
staggered dynamic susceptibility in terms of  correlations of the order
parameter. We will continue our general discussion in
Section~\ref{scalingforms} where we will present universal scaling forms for
nearly-critical coplanar  antiferromagnets.
These scaling forms follow from not much more than the presence of hyperscaling
and a dynamic critical exponent $z=1$. On the magnetically-ordered side,
the entire dynamic staggered and uniform
susceptibilities will be argued to be fully universal functions of five
parameters
characterizing the ground state:
$N_0$, the order-parameter condensate, the two stiffnesses $\rho_{\parallel}$,
$\rho_{\perp}$ and the two susceptibilities $\chi_{\parallel}$, $\chi_{\perp}$
(defined more precisely below). Similar results will hold also on the quantum
disordered side. We emphasize that none of the results of these two sections
make any specific assumptions on the universality class of the transition.

In Sections~\ref{eff_act}-\ref{qd} we will present explicit computations of
the universal scaling functions using a particular (we think likely)
field-theoretic model of the
transition. This approach has deconfined spin-1/2
spinon excitations in the quantum disordered phase, which lead to many
interesting
observable consequences. Section~\ref{applic} will compare some of the
above results with available numerical results for the $S=1/2$ triangular
Heisenberg antiferromagnet; this comparison will use some new results on
the $1/S$ expansion of this model which are obtained in Appendix~\ref{largeS}.

Our main conclusions will be reiterated in Section~\ref{conl}.
The contents of Appendix~\ref{confined} were noted above, and some technical
details will be presented in Appendix~\ref{appendt=0}.

\subsection{Order parameter and other observables}
\label{ordparm}

For simplicity, we will restrict our discussion to antiferromagnets
with Hamiltonians of the following form:
\begin{equation}
{\cal H} = \sum_{i<j} J_{ij} {\bf S}_{i} \cdot {\bf S}_{j}
\label{I1}
\end{equation}
where the ${\bf S}_i$ are spin $S$ operators on the sites
$i,j$ of a regular two-dimensional lattice,
and the $J_{ij}$ are the exchange integrals. The $J_{ij}$ respect the
symmetries of the lattice, and are short-ranged, although not necessarily
nearest neighbor.
The strength of the
quantum
fluctuations will depend on the value of $S$ and on the ratios of
the $J_{ij}$, and will determine whether the ground state is magnetically
ordered or quantum disordered.

In the following, it will be convenient to think of the ${\bf S}_i$ not
as quantum operators, but as spacetime-dependent fields in a
path integral over imaginary time $\tau$.
We will restrict our analysis to antiferromagnets in which the
strongest fluctuations
are well described by the following hydrodynamic parametrization
\begin{equation}
{\bf S}_i (\tau)  = {\bf n}_1 ({\bf x}_i, \tau) \cos(2 {\bf Q \cdot x}_i) +
 {\bf n}_2 ({\bf x}_i, \tau) \sin(2 {\bf Q \cdot x}_i )
\label{I2}
\end{equation}
where ${\bf n}_1$, ${\bf n}_2$ vary slowly on the scale of a lattice
spacing, but are always orthogonal:
${\bf n}_1~\cdot~{\bf n}_2 =0$ for all ${\bf x}_i$, $\tau$.
The ordering
wavevector $2 {\bf Q}$ may be commensurate or incommensurate with
reciprocal lattice vectors, but must not be such that (\ref{I2}) makes
all the ${\bf S}_i$ collinear with each other. Thus the square lattice
with $2 {\bf Q} = (\pi, \pi)/a$ is excluded ($a$ is the nearest neighbor
spacing), as is any ferromagnetic state (${\bf Q}=0$). The triangular lattice
and certain kagome lattice antiferromagnets with $2 {\bf Q} =
(8 \pi /3 , 8 \pi/\sqrt{3})/a$, or square lattice
antiferromagnets with incommensurate
${\bf Q}$ are however included. Kagome antiferromagnets with more complicated
local correlations, which are nevertheless coplanar, will also be described
by our universal results, but are not considered explicitly for simplicity.
The parametrization (\ref{I2}) also implies
that the spin orientations are always locally coplanar.
In fact, even antiferromagnets with non-coplanar correlations can be
analyzed by a straightforward extension (not described here)
of our results. The key restriction is that the correlations
are non-collinear: we however assume coplanarity for simplicity.

As is well-known~\cite{AM,halpsas,Dombre_Read}, we can identify the pair of
vectors ${\bf n}_1$, ${\bf n}_2$ as the order parameter of the
antiferromagnet; below we will discuss an equivalent complex matrix order
parameter, $Q_{\alpha, \beta}$,
 which is computationally somewhat more convenient.   On the
magnetically ordered there will be a spin-condensate which, we assume,
satisfies \begin{equation}
N_0^2 =  \langle {\bf n}_1 \rangle_{T=0}^2 =
 ~\langle {\bf n}_2  \rangle_{T=0}^2
\label{I3}
\end{equation}

Our analysis will rely heavily on a spinor parametrization of
the vectors ${\bf n}_1$, ${\bf n}_2$. This is most directly introduced
by the Schwinger boson representation of the spin operators
\begin{equation}
S_a = \frac{1}{2} b_{\alpha}^{\dagger} \sigma^a_{\alpha\beta} b_{\beta}
\label{II1}
\end{equation}
where $a=x,y,z$, ~$\alpha,\beta = 1,2$, and
the $\sigma^a$ are the Pauli matrices; site and time dependence
of the fields is implicit.
It turns out that the hydrodynamic form (\ref{I2}) is related to the
following parametrization of the $b$:
\begin{equation}
b_{\alpha i} (\tau ) = \sqrt{\frac{S Z_S}{2}} \left(
z_{\alpha} ( {\bf x}_i , \tau) e^{i {\bf Q} \cdot {\bf x}_i} +
i \varepsilon_{\alpha\beta}
z_{\beta}^{\ast} ( {\bf x}_i , \tau) e^{-i {\bf Q} \cdot {\bf x}_i} \right)
\label{II2}
\end{equation}
where $\varepsilon$ is the antisymmetric tensor and the $z_{\alpha}$
are slowly varying complex fields satisfying the
following normalization at some
scale $\Lambda$:
\begin{equation}
\sum_{\alpha=1}^{N} |z_{\alpha}|^2 = N
\label{zmag}
\end{equation}
with $N=2$ (we have introduced the variable
$N$ in anticipation of the generalization
below to arbitrary $N$).
The renormalization factor $Z_S$ accounts for the fluctuations at scales
shorter than $\Lambda$.
Inserting (\ref{II2}) in (\ref{II1}) and comparing with (\ref{I2}) we
obtain
\begin{equation}
n_{2a} + i n_{1a} = \frac{SZ_S}{2}  \varepsilon_{\alpha\gamma}
z_{\gamma} \sigma^{a}_{\alpha\beta} z_{\beta}
\label{II3}
\end{equation}
It is easy to check that this satisfies ${\bf n}_1 \cdot {\bf n}_2 = 0$
and (\ref{I3}). Notice that order parameters fields are {\it quadratic} in $z$,
this is consistent with the identification of the $z$ quanta as $S=1/2$ bosonic
spinons. The composite character of the order parameter was
also noticed (for $N=2$) in
 Ref.~\cite{Aza}.

Some key properties of the above parametrization deserve notice.
First, the question of gauge invariance. As is well-known,
the Schwinger boson decomposition (\ref{II1}) demands that the physics
be invariant under the $U(1)$ gauge transformation $b \rightarrow
e^{i\phi} b$. However the continuum parametrization (\ref{II2})
`breaks' this gauge symmetry~\cite{Sach-Read}. Alternatively stated, if the
$z$ fields are slowly varying in one particular choice of gauge for the $b$,
they will have forbidden rapid variations for most other gauges.
Thus, simply by focusing
on a long-wavelength theory of the $z$, we have `broken' the gauge symmetry.
There is however, a remnant $Z_2$ gauge symmetry~\cite{Sach-Read} that must
be kept track of: notice that the transformation
\begin{equation}
z({\bf x}, \tau ) \rightarrow \eta({\bf x}, \tau) z({\bf x} , \tau)
{}~~~~~~:~~~~~~\eta = \pm 1
\label{II3'}
\end{equation}
leaves all the spin operators invariant. All observables must be invariant
under this $Z_2$ gauge transformation.
All of these features are consistent with earlier large $N$ theories
of frustrated antiferromagnets~\cite{Sach-Read}
 which found breaking of $U(1)$ gauge
invariance down to $Z_2$ in all non-collinear antiferromagnets.

Consider, next, the symmetries any effective action for the $z$ must satisfy.
It must clearly be invariant under any global $SU(2)$ spin rotation
$z \rightarrow U z$, where $U$ is an $SU(2)$ matrix. More interesting,
however, is the behavior under lattice translations~\cite{Aza}, ${\bf x}
\rightarrow {\bf x} + {\bf y}$. The spin-rotation invariance of ${\cal H}$
and the parametrization (\ref{II2}) are consistent with this only if
the action is invariant under the global transformation
\begin{equation}
z \rightarrow e^{-i {\bf Q} \cdot {\bf y}} z
\end{equation}
where ${\bf y}$ is any near-neighbor vector. For the triangular lattice,
this demands that the action be invariant under the $Z_3$
symmetry~\cite{Aza} $z \rightarrow
\exp(\pm i 2 \pi /3 ) z$, while for incommensurate spiral states it
is effectively equivalent to a global $U(1)$ symmetry.
In practice we will find that the consequences of the $Z_3$ symmetry are
essentially identical to the larger $U(1)$ symmetry, and we will therefore
simply refer to this lattice symmetry as a $U(1)$ symmetry.
It is important, however,
not to confuse this global, lattice, $U(1)$ symmetry, with the
$U(1)$ gauge symmetry
discussed above. Thus the effective action for the $z$ field should
possess a global $SU(2)\times U(1)$ symmetry~\cite{Aza}; for general $N$
this will be a $SU(N)\times U(1)$ symmetry.

An important observable which will characterize the antiferromagnet,
is the staggered dynamic susceptibility $\chi_s$ defined by
\begin{equation}
\chi_s (k , i \omega_n ) \delta_{ab}
= \frac{v_s}{N_s\hbar} \sum_{i,j}
\int_{0}^{\hbar/k_B T} d \tau \left\langle S_{ia} (\tau) S_{jb} ( 0 )
\right\rangle\exp\left[-i\left(
({\bf k} + 2 {\bf Q})\cdot ({\bf x}_i - {\bf x}_j) - \omega_n \tau
\right)\right]
\label{II4}
\end{equation}
at the small momentum ${\bf k}$ away from $2 {\bf Q}$ and Matsubara
frequency $\omega_n$. The sums over $i,j$ extend over all
the $N_s$ sites of the
system, and $v_s$ is the volume per spin (e.g.,
$v_s = a^2 \sqrt{3}/2$ for triangular
antiferromagnet).

 The physically measurable
retarded staggered susceptibility can of course be obtained by the usual
analytic continuation to real frequencies.
Inserting (\ref{II1}) and (\ref{II2}) in (\ref{II4}), we find
\begin{equation}
\chi_s (k, i\omega_n ) = \frac{1}{N(N+1)\hbar}\sum_{\alpha,\beta=1}^{N}
\int d^2 x \int_0^{\hbar/k_B T} d \tau
\left\langle Q_{\alpha\beta} ( {\bf x}, \tau ) Q^{\ast}_{\alpha\beta}
(0,0) \right\rangle e^{-i({\bf k}\cdot {\bf x} - \omega_n \tau )}
\label{II5}
\end{equation}
where $N=2$, and the symmetric order-parameter $Q_{\alpha\beta}=
Q_{\beta\alpha}$ is given by
\begin{equation}
Q_{\alpha\beta} = \frac{S Z_S}{N} z_{\alpha} z_{\beta}.
\label{II6}
\end{equation}
Note that it has $N(N+1)/2$ different complex components, and
is invariant under the $Z_2$ gauge transformation (\ref{II3'}).
It transforms under $SU(N)\times U(1)$ as a $1\times 2$ Young
tableau under $SU(N)$ and as charge 2 under $U(1)$.
Again we
have introduced an $N$-dependent notation to facilitate the
generalization to arbitrary $N$. The equation (\ref{I3})
for the magnitude of the
order parameter can also be expressed in the general form
\begin{equation}
N_0^2 = \sum_{\alpha\beta=1}^{N} \left| \left\langle
Q_{\alpha\beta} \right\rangle_{T=0} \right|^2
\end{equation}
We also quote for reference the relationship, special to $N=2$, between
the tensor order parameter $Q_{\alpha\beta}$ and the vectors
${\bf n}_1$, ${\bf n}_2$, which can be deduced from (\ref{II3}) and
(\ref{II6}):
\begin{equation}
Q = \frac{1}{2} \left(
\begin{array}{cc}
-n_{2x} + i n_{2y} - i n_{1x} - n_{1y} & n_{2z} + i n_{1z} \\
n_{2z} + i n_{1z} & n_{2x} + i n_{2y} + i n_{1x} - n_{1y}
\end{array} \right)
\end{equation}

We will find it convenient to express many of our results
in terms of
the dynamic, staggered, structure factor which is the Fourier transform of the
spin-spin correlation function in real time $t$
\begin{equation}
S(k ,\omega ) ~\delta_{l,m} = \int d^2 x \int^{\infty}_{-\infty}
 d t \langle S_{l}({\bf x}, t) S_{m} (0, 0)\rangle
\exp{-i({\bf k x} - \omega t)}
\label{I7}
\end{equation}
This is of course related to the staggered susceptibility defined above
by
\begin{equation}
S(k, \omega ) = \frac{2 \hbar}{1 - e^{-\hbar \omega/(k_{B}T)}}~
{\mbox Im} \chi_s (k, \omega)
\label{I8}
\end{equation}

In addition to the order parameter, the uniform magnetization density,
${\bf M} ({\bf x})$, is an important hydrodynamic variable.
Its fluctuations decay slowly
due to the conservation law for the total magnetization.
It is defined by
\begin{equation}
{\bf M} ({\bf x}_i, t) = \frac{g \mu_{B}}{v_s} {\bf S}_i (t),
\label{I5}
\end{equation}
where  $g \mu_B/ \hbar$ is the hydromagnetic ratio.
Its diffusion is measured by the uniform spin susceptibility defined by
\begin{equation}
\chi_u (k, \omega) ~\delta_{ab} = - \frac{i}{\hbar} \int d^2 x
\int^{\infty}_{0} dt \langle [M_{a} ({\bf x}, t), M_{b} (0,0)]\rangle
\exp{-i({\bf k x} - \omega t)}
\label{I6}
\end{equation}

\subsection{Scaling forms}
\label{scalingforms}
We will now consider the properties of the non-collinear
antiferromagnets in the vicinity of a second-order
quantum phase transition from a magnetically ordered to a quantum
disordered ground state. We will try to keep the discussion in this
section as general as possible, independent of any specific field
theory for the transition. The results of this
subsection will follow from some fairly general scaling assumptions,
rather similar to those applied to collinear antiferromagnets in
Ref~\cite{CSY}. A primary assumption will be that the quantum
transition has dynamic critical exponent $z=1$.
Explicit computations of the scaling functions and exponents will be
performed in the subsequent sections using a particular deconfined-spinon
field-theory of the transition.
A confined-spinon field-theory will be briefly considered
in Appendix~\ref{confined}; its properties are also consistent with the scaling
ideas of this section.

Let us assume that the $T=0$ transition occurs as some coupling
constant $g$ is varied through a critical value $g=g_c$, and the
magnetically ordered state occurs for $g<g_c$.

We present first the scaling properties for $g<g_c$. We expect that
the condensate $N_0$ will vanish as
\begin{equation}
N_0 \sim (g_c - g)^{\bar \beta}
\label{I4}
\end{equation}
where ${\bar \beta}$ is a universal critical exponent.
A second characterization of the ordered ground state is provided by
the spin stiffnesses $\rho_{\parallel}$ and $\rho_{\perp}$: these
measure the energy cost of twists in the plane and perpendicular to
the plane of the order parameter, respectively. In the presence of
hyperscaling (which we assume), we expect that both these stiffnesses
will vanish as
\begin{equation}
\rho_{\perp}, \rho_{\parallel} \sim (g_c - g )^{\nu}
\label{I4'}
\end{equation}
where the $\nu$ is the usual correlation length exponent (this
formula is special to two dimensions).
Further, the ratio of these two stiffnesses will obey
\begin{equation}
\lim_{g\nearrow g_c} \frac{\rho_{\parallel}}{\rho_{\perp}} = \Upsilon_{\rho}
\label{defups}
\end{equation}
where $\Upsilon_{\rho}$ is a {\em universal\/} number.
In a similar manner we can consider the two uniform magnetic susceptibilities
$\chi_{\parallel}$, $\chi_{\perp}$ defining the response of the antiferromagnet
with infinitesimal anisotropy to uniform magnetic fields perpendicular
and parallel to the plane of the order parameter, respectively (note the
inversion in the order of `parallel' and `perpendicular' !).
In a $z=1$ theory their scaling
properties are identical to those of the spin stiffnesses, and possess
an associated universal ratio $\Upsilon_{\chi}$. The subsequent sections
of this paper consider a field theory in which $\Upsilon_{\rho} =
\Upsilon_{\chi} = 1$ exactly; in Appendix~\ref{confined}
we briefly consider a model
with different universal ratios. In all cases it is useful to
consider the dimensionless numbers $y_{\rho}$, $y_{\chi}$
\begin{equation}
y_{\rho} = \frac{\rho_{\parallel} - \Upsilon_{\rho}
\rho_{\perp}}{\rho_{\perp}}~~~~;~~~~
y_{\chi} = \frac{\chi_{\parallel} - \Upsilon_{\chi}
\chi_{\perp}}{\chi_{\perp}}
\end{equation}
which measure the deviation of the stiffnesses and susceptibilities from
the universal ratio at the critical point; clearly $y_{\rho}, y_{\chi}
\rightarrow 0$ as $g \rightarrow g_c$.
Finally, as in Ref~\cite{CSY}, we also need the following dimensionless
ratios which measure the wavevector, frequency, and stiffness in
units of the absolute temperature
\begin{equation}
\overline{k} = \frac{\hbar c_{\perp} k}{k_B T} ~~;~~
\overline{\omega} = \frac{\hbar \omega}{k_B T}~~;~~
x_1 = \frac{N k_B T}{4 \pi \rho_{\perp}};
\end{equation}
The numerical factor of $4 \pi$ is for future notational convenience,
and the spin-wave velocities $c_{\perp}$, $c_{\parallel}$ are of
course given by
$c_{\perp}^2 = \rho_{\perp}/ \chi_{\perp}$ and $c_{\parallel}^2 =
\rho_{\parallel}/ \chi_{\parallel}$. The factor $N$ in $x_1$ has
been inserted because $\rho_{\perp} \propto N$ in the large limit,
and so ensures that $x_1$ remains of order unity in this limit.

Now, following arguments closely related to those in Ref.~\cite{CSY},
we may conclude that the response functions of nearly-critical antiferromagnets
obey the following universal scaling forms
\begin{eqnarray}
\chi_s (k ,\omega) &=& \frac{2 \pi N^{2}_0}{N \rho_{\perp}}~
\left(\frac{N k_B T}{4 \pi \rho_{\perp} }\right)^{\bar \eta}~
 \left(\frac{\hbar c_{\perp}}{k_B T}\right)^2 ~
\Phi_{1s} \left(\overline{k}, \overline{\omega}, x_1, y_{\rho}, y_{\chi}
\right) \\
\chi_u (k ,\omega) &=& \left(\frac{g \mu_B}{\hbar c^{2}_{\perp}}\right)^2
{}~k_B T ~\Phi_{1u} \left(\overline{k},
\overline{\omega}, x_1, y_{\rho}, y_{\chi} \right) \\
S (k ,\omega) &=& ~\frac{2 \pi \hbar N^{2}_0}{N \rho_{\perp}}~
\left(\frac{N k_B T}{4 \pi \rho_{\perp}}\right)^{\bar \eta}~
\left(\frac{\hbar c_{\perp}}{k_B T}\right)^2 ~
{}~\frac{2}{1 - e^{-\overline{\omega}}}~
\Xi_1 \left(\overline{k}, \overline{\omega}, x_1, y_{\rho}, y_{\chi}\right)
\label{I30}
\end{eqnarray}
Here
$\Phi_{1s}$,  $\Phi_{1u}$ and $\Xi_1$
are completely universal functions of their dimensionless arguments
and there are no non-universal scale factors anywhere. The exponent
$\bar \eta$ is related to the order parameter exponent $\bar \beta$
by the hyperscaling relation
\begin{equation}
2 \bar \beta = ( 1 + \bar \eta ) \nu .
\end{equation}
{}From the above scaling relation and (\ref{I4}) and (\ref{I4'}) we see
that the prefactors of all the scaling functions remain finite all the way
up-to $g=g_c$, or $x_1 = \infty$. Further,
all scaling functions are defined such that they
remain finite as $x_1 \rightarrow \infty$ when we will also find
$y_{\rho,\chi}
\rightarrow 0$. The universal functions
$\Phi_{1s}$ and $\Xi_1$ are related by the fluctuation-dissipation
 theorem $\Xi_1  =
\mbox{Im} \Phi_{1s} $. As in~\cite{CSY},
the argument $x_1$ determines whether the system is better described at large
scales by a quantum-critical ($x_1 \gg 1$) or a renormalized-classical
($x_1 \ll 1$) theory.

Strictly speaking, the leading scaling properties of the observables are
obtained at $y_{\rho} = y_{\chi} = 0$, because these ratios are associated
with irrelevant operators. However many long-distance properties are
sensitive to the precise values of the spin-stiffnesses and susceptibilities.
Thus these operators are actually {\em dangerously irrelevant\/},
and it necessary to consider many observables as full functions of
$y_{\rho}$ and $y_{\chi}$

Parallel arguments can be applied to the quantum disordered state
with $g > g_c$. We assume that this state has low-lying quasiparticle
excitations with non-zero spin,
characterized by an energy scale $\Delta$, which propagate
with a velocity $c$. In the model considered in the subsequent sections
we will have spin-1/2, bosonic quasiparticles above a gap $\Delta$;
there are however other possibilities, one of which is discussed
in Appendix~\ref{confined}.
We expect that $\Delta$ will obey
\begin{equation}
\Delta \sim (g - g_c)^{\nu}
\label{deltanu}
\end{equation}
near the critical point. We also introduce the dimensionless ratio
\begin{equation}
x_2 = \frac{k_B T}{\Delta}
\end{equation}
which is the analog of the $x_1$ on the ordered side. There is now no need
to consider the analogs of the $y_{\rho}$, $y_{\chi}$ as these
will be truly irrelevant (as opposed to dangerously irrelevant) on the
disordered side. The observables of the nearly-critical, quantum-disordered
antiferromagnet obey
\begin{eqnarray}
\chi_s (k ,\omega) &=& {\cal A} ~\left(\frac{\hbar c}{k_B T}\right)^2
{}~\left(\frac{ k_B T}{\Delta}\right)^{\bar \eta}~
\Phi_{2s} \left(\overline{k}, \overline{\omega}, x_2 \right) \\
\chi_u (k ,\omega) &=& \left(\frac{g \mu_B}{\hbar c^{2}}\right)^2
{}~k_B T ~\Phi_{2u} \left(\overline{k}, \overline{\omega},
 x_2 \right) \\
S (k ,\omega) &=& \hbar {\cal A}
{}~\left(\frac{\hbar c}{k_B T}\right)^2 ~
\left(\frac{k_B T}{\Delta}\right)^{\bar \eta}~
{}~\frac{2}{1 - e^{-\overline{\omega}}}~
\Xi_2 \left(\overline{k}, \overline{\omega}, x_2 \right)
\label{extr}
\end{eqnarray}
Again $\Phi_{2s}$, $\Phi_{2u}$ and $\Xi_2$ are completely universal functions.
The prefactor ${\cal A}$ is related to quasiparticle amplitude(s)
and vanishes as
\begin{equation}
{\cal A} \sim (g - g_c)^{\bar \eta \nu}
\end{equation}
The precise definition of ${\cal A}$ requires a normalization condition
on $\Phi_{2s}$ which will be discussed later.

Before closing this section, we briefly introduce the scaling functions
of some other important observables which can be deduced from the
ones above.
We restrict
ourselves to the ordered side; the extension to the
disordered side is straightforward.
The scaling function for the spin correlation length is
\begin{equation}
\xi^{-1} = \frac{k_B T}{\hbar c_{\perp}} ~X (x_1, y_{\rho}, y_{\chi})
\label{I33}
\end{equation}
The static uniform spin susceptibility at $g < g_c$ behaves as
\begin{equation}
\chi_u (T) = \left (\frac{g \mu_B}{\hbar c_{\perp}}\right)^2 k_B T
{}~\Omega (x_1,  y_{\rho}, y_{\chi})
\label{I34}
\end{equation}
The local structure factor  $S_L (\omega)$ is given by
 $ S_L (\omega) =
\int d^2 k ~S (k, \omega)/ 4 \pi^2$. The contribution
of $\chi_u$ to $S_L (\omega)$ is subdominant and $S_L$ is given simply
by a momentum integral of the staggered susceptibility.
This integral
is always ultraviolet convergent (because the intermediate states in
$S(k,\omega)$ are all on-shell) and is dominated by
 $\overline{k}$ less than about 1; we have therefore
\begin{equation}
S_L (\overline{\omega}) = \frac{2 \pi \hbar N^{2}_0}{N \rho_{\perp}}
 ~~\left(\frac{N k_B T}{4
\pi \rho_{\perp}}\right)^{\bar \eta} ~\frac{2}{1 - e^{-\overline{\omega}}}
{}~K_1 (\overline{\omega}, x_1,  y_{\rho}, y_{\chi} )
\label{I35}
\end{equation}
where $K_1  =
 \int d^2 \overline{k} ~\Xi_1 /4 \pi^2$.
The small frequency limit of $S_L (\omega)$ is directly related to the
spin-lattice relaxation rate $1/T_1 \propto
S_L(\overline{\omega} \rightarrow 0)$.
We will also discuss static structure factor
$S(k) = \int d \omega  ~S (k, \omega)~/2\pi$. The frequency integral is
divergent at the upper cutoff if ${\bar \eta} > 1$, whence $S(k)$ is
non-universal - this will be the case in our model.

In the subsequent sections we will obtain explicit expressions for the scaling
functions introduced above in the renormalized-classical
and quantum-critical regions.
We will use a new deconfined spinon field theory which will be introduced in
Section~\ref{eff_act}, along with a $1/N$ expansion which will facilitate our
computations.
Properties of the quantum-disordered phase will also be discussed.

\section{Effective field theory: deconfined spinons}
\label{eff_act}

The main aim of the remainder of this paper is to illustrate the
general scaling ideas discussed above in the framework of a specific
field theoretical model of the quantum transition.
An important property of field-theory we use is that results of the
$1/N$, spacetime $D=4-\epsilon$, and $D = 2 + \epsilon$, expansions
on it are {\em all\/} consistent with each other; we will consider
only the $1/N$ expansion here.

A significant reason behind the choice of our particular model
is that it possesses deconfined spin-1/2 spinon excitations in the quantum
disordered state. This is then
consistent with the $Sp(N)$-large $N$
prediction of Ref.~\cite{Sach-Read} on
non-collinear antiferromagnets. Further, our approach will allow us to explore
some of the observable consequences of these novel excitations.

We begin with some discussion on the role of the $Z_2$ gauge symmetry of
(\ref{II3'}). The crucial role of this gauge symmetry was noted in
was emphasized to us
at an early stage by N. Read~\cite{Nick} and was also
noted in Ref~\cite{Sach-Read}.
Our main assumption will be that the $Z_2$
gauge symmetry can be entirely neglected in the continuum field theory.
In other words, configurations with a non-zero local $Z_2$ flux
remain gapful across the transition.
The $Z_2$ gauge fluxes are in fact present in the cores of vortex lines
(in spacetime)
associated with homotopy group $\pi_1 ( SO(3)) = Z_2$
of the true $SO(3)$ order parameter~\cite{Kaw-Miyashita}.
We assume that these vortices remain confined across the transition
and that the $Z_2$ gauge charge of the $z$ field is globally
defined~\cite{Nick} (the $z$-field configuration around a vortex is
double-valued). Under these circumstances we may simply write down a
continuum  Landau-Ginzburg field theory for the $z$-field.
Implicitly, this procedure implies that we are not distinguishing
between $SU(2)$ and $SO(3)$ symmetries.

We will now write down the most general action consistent with
the $SU(N)\times U(1)$ symmetry discussed before. Rather than using
a soft-spin Landau-Ginzburg approach, we find it more
convenient to use hard spins satisfying (\ref{zmag}); this modification
is however not crucial and completely equivalent results can be obtained
by the former approach. To second-order in spatial gradients this yields
the following effective action
\begin{equation}
{\cal S} =
\int d^2 x d \tau \sum_{\mu = \vec{x}, \tau} \frac{1}{g_{\mu}}
\left[ \partial_{\mu} z^{\ast}_{\alpha} \partial_{\mu} z_{\alpha}
- \frac{\gamma_{\mu}}{4 N}
\left( z^{\ast}_{\alpha} \partial_{\mu} z_{\alpha} - \partial_{\mu}
z^{\ast}_{\alpha} ~z_{\alpha} \right)^2 \right].
\label{C1}
\end{equation}
where $\alpha = 1 \ldots N$, and
$g_x$, $g_{\tau}$, $\gamma_x$, $\gamma_{\tau}$ are coupling constants.
Any of these couplings can be varied to tune through
the quantum transition - we will use
\begin{equation}
g \equiv g_x.
\end{equation}
Simple considerations presented in Section~\ref{cons_curr} below show that
these
coupling constants are given by
\begin{equation}
g_x = \frac{N}{2 \rho^{0}_{\perp}}, ~~~g_{\tau} =
\frac{N}{2 \chi^{0}_{\perp}},
{}~~~\gamma_x = \frac{\rho^{0}_{\parallel} -
 \rho^{0}_{\perp}}{\rho^{0}_{\perp}}, ~~~
{}~~~\gamma_{\tau} = \frac{\chi^{0}_{\parallel} -
\chi^{0}_{\perp}}{\chi^{0}_{\perp}},
\label{C1'}
\end{equation}
and $\rho^0$ and $\chi^0$ are the bare values of two spin stiffnesses and spin
susceptibilities. For simplicity,
 throughout the paper  we define transverse and longitudinal susceptibility
 without a factor $g \mu_B /\hbar$.
A more detailed consideration of the values of $\rho$, $\chi$ is presented
in the next section.

The effective action ${\cal S}$ can also be explicitly derived
from microscopic considerations. Using the continuum parametrization in
(\ref{II2}) it is not difficult to show that the long-distance limit
of the $Sp(N)$ theories of
Refs~\cite{Sach-Read} and~\cite{Kag} is described precisely by ${\cal S}$.
The same parametrization can also be used on the semiclassical approach
of~\cite{Dombre_Read} to obtain ${\cal S}$. Finally,
 we also explicitly derived the effective action
(\ref {C1}) for $N=2$ from the general macroscopic
approach of Ref~\cite{AM}.

Some critical properties of ${\cal S}$ can be immediately deduced.
By a simple power-counting argument in $D=4-\epsilon$ dimensions it can
shown that the $\gamma_{\mu}$ couplings are irrelevant at the critical point.
An identical result can also be obtained by $D=2+\epsilon$ analysis
parallel to that of Ref~\cite{Azaria}. We will also explicitly show
the irrelevancy of the $\gamma_{\mu}$  in the $1/N$ expansion below.
(None of these arguments of course exclude the possibility that a large
bare value of $\gamma_{\mu}$ may have more fundamental effects. In fact, at
$\gamma_{\mu}=-N$, ${\cal S}$ actually becomes the $U(1)$ gauge invariant
$CP^{N-1}$ model, which is then a model for quantum phase transitions in
collinear antiferromagnets. We will not consider the possibility of these
large $\gamma_{\mu}$ complications in this paper.)

It is therefore useful to begin the analysis by considering
${\cal S}$ at $\gamma_{\mu} = 0$. It is easy to verify that now
${\cal S}$ has its internal symmetry enlarged from $SU(N)\times U(1)$
to $O(2N)$. Further, the spacetime theory is Lorentz invariant.
Finally, this theory has $\rho_{\parallel}=\rho_{\perp}$
(and similarly for $\chi$) and so we have
$\Upsilon_{\rho} = \Upsilon_{\chi} = 1~\mbox{exactly}$

The  exponents appearing in the scaling functions
are now all properties of the well-known $O(2N)$ fixed point,
and we quote for reference to order $1/N$ (see also Appendix~\ref{appendt=0})
\begin{equation}
\nu = 1 - \frac{16}{3 \pi^2 N}~;~\bar \eta = 1 + \frac{32}{3 \pi^2 N}
{}~;~\bar \beta = 1 + {\cal O} (1/N^2)
\label{exponents}
\end{equation}
Note that the exponents $\bar \eta$ and $\bar \beta$ are associated
with the composite field $Q_{\alpha\beta}$ and thus differ from the
usual $\eta$, $\beta$ for vector fields. Thus $\bar \eta$ is quite
close to unity at large $N$, while the corresponding $\eta$ which appears in
collinear antiferromagnets is almost zero.

Let us now consider how the $\gamma_{\mu}$ variables break the
Lorentz and $O(2N)$ symmetry. Tedious but straightforward computations
show that the terms proportional to the $\gamma_{\mu}$ transform under
a single irreducible representation of $O(2N)$ - the one labeled by
a Young tableau of 2 rows and 2 columns. It is therefore not necessary
to decompose the $O(2N)$ structure of the operator. However the
$\gamma_{\mu}$ terms are irreducible under the Lorentz group - there are the
spin-0 and spin-2 pieces
\begin{eqnarray}
\gamma_1 &=& (2 \gamma_{x} +  \gamma_{\tau})/3 \nonumber \\
\gamma_2 &=& (\gamma_{x} - \gamma_{\tau})/3.
\label{defgamma1}
\end{eqnarray}
The terms associated with the $\gamma_1$ and $\gamma_2$ are now completely
irreducible under $O(2N)$ and Lorentz group, and will therefore have their
independent crossover exponents, $\phi_1$ and $\phi_2$ respectively,
measuring their irrelevancy.  In other words,
near the quantum fixed point,
the fully renormalized spin stiffnesses and susceptibilities should obey
\begin{eqnarray}
\frac{\rho_{\parallel} -
\rho_{\perp}}{\rho_{\perp}} &=&  \gamma_{1} (\xi_{J})^{-\phi_1} + \gamma_2
(\xi_J)^{-\phi_2} \nonumber\\
\frac{\chi_{\parallel} -
\chi_{\perp}}{\chi_{\perp}} &=&  \gamma_{1} (\xi_{J})^{-\phi_1} - 2\gamma_2
(\xi_J)^{-\phi_2}
\label{C14}
\end{eqnarray}
where $\xi_J$ is the Josephson correlation length measured in lattice units.
To leading order in $\gamma$ we also have from (\ref{C14}) for the
spin-wave velocity difference,
\begin{equation}
\frac{c_{\parallel} - c_{\perp}}{c_{\perp}} =  \frac{3}{2}~
\gamma_{2} (\xi_{J})^{-\phi_2}
\label{C14'}
\end{equation}
As $g_{\mu}$ approaches $g^{c}_{\mu}$, $\xi_J$ behaves as
$\xi_J \sim (1 - g_{\mu}/g^{c}_{\mu})^{-\nu}$ (clearly, $g_{x}/g^{c}_{x} =
g_{\tau}/g^{c}_{\tau})$.
 In section~\ref{cons_curr} we will find
the following $1/N$ expansion result for the renormalization-group
eigenvalues attracting the $\gamma_{1,2}$ to the fixed point
\begin{equation}
\phi_1 = 1 + \frac{32}{3 \pi^2 N}~~~;~~~\phi_2 = 1 + \frac{112}{15 \pi^2 N}
\label{defphi1}
\end{equation}

\section{Conserved charges and currents}
\label{cons_curr}

This section will present the computation of the spin stiffnesses
and uniform spin susceptibilities
both at $T=0$ and in the quantum critical region of the
deconfined spinon action ${\cal S}$. The calculation will be carried
out to order $1/N$. We will show how one can obtain renormalized stiffnesses
in the ground state
by doing calculations in the symmetric phase at $T \rightarrow 0$. A
 computation of stiffnesses directly in the ordered phase is performed
in the Appendix~\ref{appendt=0}.

The stiffnesses and uniform susceptibilities are all response
functions associated with the conserved charges and currents of
${\cal S}$. We will therefore begin by studying the
$SU(N) \times U(1)$ symmetry of ${\cal S}$.
The conserved charges and currents can be determined by the usual
procedure of
evaluating the change in the action
under an infinitesimally small
symmetry transformation with a spacetime dependent angle. The results
are conveniently expressed in terms of the
$N^2 -1$ traceless Hermitian $SU(N)$ generators $T^a$
which we choose to satisfy
\begin{equation}
\mbox{Trace} \left( T^a T^b \right) = \frac{1}{2} \delta^{ab}
\label{C4}
\end{equation}
Then the currents associated with the $SU(N)$ symmetry can be written as
\begin{equation}
K_{\mu}^a = - \frac{i}{g_{\mu}} \left( z^{\dagger} T^a \partial_{\mu} z -
 \partial_{\mu}
z^{\dagger} T^a z \right) - \frac{i\gamma_{\mu}}{N g_{\mu}} \left(
z^{\dagger} \partial_{\mu} z - \partial_{\mu} z^{\dagger} z \right) \left(
z^{\dagger} T^a z \right)
\label{C3'}
\end{equation}
The index $\mu$ extends over $\vec{x}, \tau$ and $K_{\tau}^a$ is,
strictly speaking, a conserved charge density - in this section we
will use the term `current' to generically refer to both charges and currents.
We will also not explicitly display the
spacetime-dependence of the fields.
The current associated with the $U(1)$ symmetry is
\begin{equation}
J_{\mu} = -\frac{i}{g_{\mu}}~(1 + \gamma_{\mu}) \left(
z^{\dagger} \partial_{\mu} z - \partial_{\mu} z^{\dagger} z \right)
\label{C5}
\end{equation}

Our intention is to express the  fully renormalized stiffnesses in terms
of $SU(N)$ and $U(1)$ current-current correlators, and so we need the
appropriate Kubo formula. To derive this formula it is convenient to
 introduce vector potentials
which linearly couple to the conserved currents above,
and examine the response of the
system to these vector potentials.
Let us consider first a $SU(N)$ vector potential
$A^{a}_{\mu}$.  This modifies the action to
\begin{equation}
{\cal S}^{\prime} =  \frac{1}{g_{\mu}} \int d^2 x d \tau \left[ \left|
\left( \partial_{\mu} + i A_{\mu}^a T^a \right) z \right|^2
- \frac{\gamma_{\mu}}{4 N} \left( z^{\dagger}
 \left( \partial_{\mu} z + i A_{\mu}^a
T^a z\right) -  \left( \partial_{\mu} z^{\dagger} - i A_{\mu}^a
z^{\dagger} T^a \right) z \right)^2 \right]
\label{C6}
\end{equation}
It is then not difficult to obtain the response of the free energy $F = - \log
\left[ \int {\cal D} z e^{-{\cal S}^{\prime}} \right]$
to the external vector potential. Doing the algebra we find
\begin{equation}
\frac{\delta^2 F}{\delta A_{\mu}^a \delta A_{\mu}^b} = - \left\langle
K_{\mu}^a K_{\mu}^b \right\rangle_S + \frac{1}{g_{\mu}} \left\langle
z^{\dagger} \left( T^a T^b + T^b T^a \right)  z\right\rangle_S
+ \frac{2 \gamma_{\mu}}{N g_{\mu}}
\left\langle z^{\dagger} T^a z z^{\dagger} T^b z
\right\rangle_S
\label{key1}
\end{equation}
Again space-time dependences have been suppressed, and the two fields
inside the correlator are at different spacetime points.
A very similar analysis can be carried out for
a $U(1)$ vector potential $a_{\mu}$,
and we find
\begin{equation}
\frac{\delta^2 F}{\delta a_{\mu}^2} = - \left\langle
J_{\mu} J_{\mu} \right\rangle_S
+ \frac{2 N (1 + \gamma_{\mu})}{g_{\mu}}
\label{key2}
\end{equation}

Now we change tracks and  evaluate the response of the system to
these vector potentials in an entirely different way. Let us
 assume that we are on the ordered side ($g < g_c$),
 and are able to integrate out all the fluctuations, including the amplitude
fluctuation modes in the direction of the condensate. We then obtain
a {\em fully} renormalized action for the spin-wave fluctuations.
Let this effective
action have the following form
\begin{equation}
F = 2\int d^2 x d \tau \left[\rho_{1\mu} \left| \partial_{\mu} Z \right|^2
-  \left( \rho_{2\mu} - \rho_{1\mu} \right) \left(
Z^{\dagger} \partial_{\mu} Z \right)^2 \right]
\label{C7}
\end{equation}
Here $Z$ is a $N$-component complex vector of {\em unit} length which
yields the local orientation of the condensate. The factor of 2 in $F$ is
introduced for further convenience. Let the condensate point
in some fixed direction $Z_0 = (1,0,0,0,\ldots )$. We now look at small
variations about this direction as in
\begin{equation}
Z = Z_0 + (i \sigma , \pi_1 + i \pi_2,
\pi_3 + i \pi_4 , \ldots)/2;
\label{Zord}
\end{equation}
the factor of $1/2$ is present because $Z$ is a spinor and rotates by
only half the angle of the observable order parameter. The
effective action for this variation is
\begin{equation}
F = \frac{1}{2} \int d^2 x d \tau \left[ \rho_{1\mu} \sum_{i=1}^{2N-2}
\left( \partial_{\mu} \pi_i \right)^2 + \rho_{2\mu}
\left( \partial_{\mu} \sigma \right)^2 \right]
\label{C8}
\end{equation}
By the definition of the stiffnesses we
identify $\rho_{1x} , \rho_{2x}$ as the two spin stiffnesses of the
spin wave modes:
\begin{equation}
\rho_{1x} = \rho_{\perp}~~~~,~~~~\rho_{2x} = \rho_{\parallel}
\end{equation}
Also the stiffness to twists in the time direction gives us the
uniform spin susceptibility:
\begin{equation}
\rho_{1\tau} = \chi_{\perp}~~~~,~~~~ \rho_{2\tau} = \chi_{\parallel}
\end{equation}
Now let us look at the response of $F$ to the presence of
an external $SU(N)$ vector potential, while the condensate is non-zero.
Doing the same analysis as before we obtain
\begin{equation}
\delta F = 2 \int d^2 x d\tau \left[\rho_{1\mu} A_{\mu}^a A_{\mu}^b
Z_0^{\dagger} T^a T^b Z_0 +  \left( \rho_{2\mu}
- \rho_{1\mu} \right) A_{\mu}^a A_{\mu}^b \left(Z_0^{\dagger} T^a Z_0 \right)
\left(Z_0^{\dagger} T^b Z_0 \right) \right]
\label{key5}
\end{equation}
For a fixed condensate $Z_0$ this result
will  depend upon the orientation
of the $SU(N)$ rotation $A_{\mu}^a$. However if we place the system
in a box of large, but finite length $L_{\mu}$ in the $\mu$ direction
the response of $F$ is clearly proportional to
$\delta^{ab}$ because
no symmetry can be broken (for the case $\mu=\tau$ this equivalent
to having a small finite temperature $T \propto L_{\tau}^{-1}$).
Thus we should replace each $T^{a} T^b$
factor by
its average over all the generators of $SU(N)$ - it is crucial that we average
over all the generators, and not over different orientations
of the condensate.
We will then need the identities
\begin{eqnarray}
\frac{1}{N^2-1} \sum_a Z_0^{\dagger} T^a T^a Z_0 &=& \frac{1}{2N} \nonumber\\
\frac{1}{N^2-1} \sum_a \left( Z_0^{\dagger} T^a Z_0 \right)^2 &=&
\frac{1}{2N (N+1)}
\label{key6}
\end{eqnarray}
which are actually true for {\em any} complex unit vector $Z_0$.
These identities can
be easily established by considering
explicit forms for the $T^a$. So finally,
combining (\ref{key5}) and (\ref{key6}),
we can determine the response of $F$ to the $SU(N)$ vector potentials at
an infinitesimal temperature:
\begin{equation}
\frac{\delta^2 F}{\delta A_{\tau}^a \delta A_{\tau}^b} = \delta^{ab}
{}~\frac{2}{N}~\frac{N \chi_{\perp} + \chi_{\parallel}}{(N+1)}
{}~~;~~\frac{\delta^2 F}{\delta A_{x}^a \delta A_{x}^b} = \delta^{ab}
{}~\frac{2}{N}~\frac{N \rho_{\perp} + \rho_{\parallel}}{(N+1)}
\label{key3}
\end{equation}
We will evaluate the left-hand side using (\ref{key1}) and thence
obtain an expression for the above linear combination of the
stiffnesses.

We still need a second linear combination - this is of course
provided by the $U(1)$ currents.
An exactly parallel computation can be done for the response to the $U(1)$
vector potential $a_{\tau}$ - in this case we find
\begin{equation}
\frac{\delta^2 F}{\delta a_{\tau}^2} = 4 \chi_{\parallel}~~;~~
\frac{\delta^2 F}{\delta a_{x}^2} = 4 \rho_{\parallel}
\label{key4}
\end{equation}
Combined with (\ref{key2}), (\ref{key1}) and (\ref{key2}) we now have
reduced determination of the spin stiffnesses and susceptibilities to
evaluation of the correlators in (\ref{key1}), (\ref{key2}) at $T=0$.
This calculation will be carried out in Section~\ref{t=0stiff} to
order $1/N$.

These methods can also be used to obtain the temperature dependence
of the uniform spin susceptibility $\chi_u (T)$. By analysis similar
to that in Ref~\cite{CSY}
it is not difficult to show that
\begin{equation}
\chi_u (T) = \left(\frac{g \mu_B}{\hbar}\right )^2 ~
\frac{\delta^2 F}{\delta A_{\tau}^a \delta A_{\tau}^a}
\label{C9}
\end{equation}
where there is no summation over $a$.
This computation will be considered in Section~\ref{unifchi}.

We conclude with a note on the nature of the $1/N$ expansion
of ${\cal S}$.
We found that a properly renormalized theory
for the scaling functions can only be defined if we restrict with the leading
terms in an expansion in powers of $\gamma_{\mu}$:
we shall therefore do a double expansion
in powers of $1/N$ and $\gamma_{\mu}$. This expansion is most easily done be
treating the effects of $\gamma_{\mu}$ perturbatively - {\em i.e.} without
introducing a Hubbard - Stratonovich decoupling of the quartic term.

\subsection{Spin-stiffnesses and susceptibilities at $T=0$}
\label{t=0stiff}

Below we will need the form for the vertex function associated with the
 anisotropic term in the action. In the  momentum space we have
\begin{equation}
\Gamma_{\alpha ,\beta} (k_{1,\mu}, k_{2,\mu}; k_{3,\mu}, k_{4,\mu}) =
\frac{\gamma_{\mu}}{4
N}~z^{\dagger}_{1,\alpha}z^{\dagger}_{2,\beta}z_{3,\alpha}
z_{4,\beta}~(k_{1,\mu} + k_{3,\mu})(k_{2,\mu} + k_{4,\mu})
\label{C10}
\end{equation}
where $\alpha$ and $\beta$ number the components of the $z$-field.
 The diagrammatic representation for
the current-current correlation functions is shown in Fig.\ref{figcurrent}.
At $N=\infty$, one
can neglect self-energy and vertex correction within a bubble; however the
renormalization  due to $\Gamma$ generally cannot be
neglected because the  summation over the components of the $z$-field in
the extra bubble associated with $\Gamma$ yields a factor of $N$ which
cancels out the $1/N$ factor in (\ref{C10}). However, a simple inspection of
the diagrams shows that the effects of $\Gamma$ are relevant at $N = \infty$
only for the $U(1)$ correlator, while for the $SU(N)$ currents, the side
vertices in the
bubble  contain sign-oscillating  $T$ matrices,
 and the  summation over the components of $z-$ field gives only
a factor ${\cal O}(1)$. As a result, we find at $N \rightarrow \infty$ and
 in the limit $T \rightarrow 0$
\begin{equation}
2N \frac{\delta^2 F}{\delta A_{\mu}^a \delta A_{\mu}^b} = 2N \delta^{ab}
\left( \frac{1}{g_{\mu}} - \frac{1}{g^{c}_{\mu}} \right)
\label{C11}
\end{equation}
and
\begin{equation}
\frac{\delta^2 F}{\delta a_{\mu}^2} = 2N \left( \frac{1}{g_{\mu}} -
 \frac{1}{g^{c}_{\mu}}
\right) \left[ 1 + \gamma_{\mu} \frac{(g^{c}_{\mu} - g_{\mu})}{g^{c}_{\mu}}
 \right]
\label{C12}
\end{equation}
where $g^{c}_{x} = g_c$ and $g^{c}_{\tau} = c^{2}_{\perp} g_{c}$.
Clearly from (\ref{key3}, \ref{key4}),
the r.h.s. in (\ref{C11}) and (\ref{C12})
 are  also the values of $4 \rho_{\perp\mu}$
and $4 \rho_{\parallel\mu}$
respectively. Note that, as one might expect,
only $\rho_{\parallel}$ acquires a correction
due to $\gamma$, while $\rho_{\perp}$ remains the same as in the isotropic
case. Also note that (\ref{C12}) is indeed consistent with (\ref{C14}) and
establishes that $\phi_1 = \phi_2 =1$ at $N = \infty$.

We now describe the $1/N$ corrections. Obviously,
 we have to consider the self-energy and vertex corrections within a
bubble in Fig.\ref{figcurrent},
and the renormalization of the vertex function $\Gamma$ itself. The
latter however is again relevant only for $U(1)$ response, while
 for $SU(N)$ response, the leading effect of $\gamma_{\mu}$
is itself of order $\gamma_{\mu}/N$ and there is no need to consider
the renormalization of $\Gamma$ to order $1/N$. The computation of the
$SU(N)$ response  therefore requires less efforts, and evaluating
the diagrams in Fig.\ref{figcurrent} with  $\Gamma$ given by (\ref{C10}),
 we obtain
\begin{equation}
2N \frac{\delta^2 F}{\delta A_{\mu}^a \delta A_{\mu}^b} = 2N \delta^{ab}
\frac{1 + \gamma_{\mu} /(2N)}{g_{\mu}}~
\left[ \left(1 - \frac{g_{\mu}}{\bar{g}^{c}_{\mu}}\right)^{\nu} +
\frac{\gamma_{\mu}}{2N}~
\left(1 - \frac{g_{\mu}}{\bar{g}^{c}_{\mu}}\right)^{2\nu}
 \right]
\label{C15}
\end{equation}
where $\bar{g}^{c}_{\mu} = g^{c}_{\mu} (1 + \gamma_{\mu}/2N)$ and $\nu = 1 -
16/3 \pi^2 N$ is the critical exponent for the correlation length.

Our next step will be to calculate, with logarithmic accuracy,
the renormalized value of $\Gamma$ as $T \rightarrow 0$.
We will then use the result to compute the $U(1)$ response to order $1/N$.
 The diagrams
which contribute to the vertex renormalization to order $1/N$
are shown in Fig. \ref{figvertren}. The internal part of each diagram
contains two Green functions and the polarization operator - this combination
produces logarithms after {\em integration}
over intermediate momentum and frequency
in $2+1$ dimensions~\cite{CSY}. The evaluation of diagrams
 is tedious but straightforward, and after doing the  algebra we obtained
that the momentum dependence of the vertex remains the same as in (\ref{C10})
but $\gamma_{\mu}$ changes to $\gamma^{eff}_{\mu}$ where
\begin{eqnarray}
\gamma^{eff}_{\tau} &=& \gamma_{\tau} \left(1
 + \frac{128}{15 \pi^2 N} \log(1-g_x/g_c)\right) +
\gamma_{x}~\frac{32}{15 \pi^2 N} \log(1-g_x/g_c)
 \nonumber \\
\gamma^{eff}_{x} &=&
\gamma_{x} \left(1
 + \frac{48}{5 \pi^2 N}\log(1-g_x/g_c)\right)
+ \gamma_{\tau}~\frac{16}{15 \pi^2 N}\log(1-g_x/g_c)
\label{C16}
\end{eqnarray}
Substituting the renormalized vertex into the bubble diagram for $U(1)$
response and performing also self-energy and vertex renormalizations within
each bubble in the way  described in~\cite{CSY}, we obtain to order $1/N$
\begin{equation}
\frac{\delta^2 F}{\delta a_{\mu}^2} \equiv 4 \rho_{\parallel ,\mu} =
 \frac{2N}{g_{\mu}}~ \left(1 + \frac{\gamma_{\mu}}{2N} \right)~ \left[
\left(1- \frac{g_{\mu}}{\bar{g}^{c}_{\mu}} \right)^{\nu} + \gamma^{eff}_{\mu}
\left(1 +  \frac{1}{2N}\right)~\left(1-
\frac{g_{\mu}}{\bar{g}^{c}_{\mu}} \right)^{2\nu} \right].
\label{C17}
\end{equation}
Eqns (\ref{C15}) and (\ref{C17}) can now be combined to obtain transverse
stiffness to order $1/N$
\begin{equation}
4 \rho_{\perp, \mu} =
 \frac{2N}{g_{\mu}}~ \left(1 + \frac{\gamma_{\mu}}{2N} \right)~ \left[
\left(1- \frac{g_{\mu}}{\bar{g}^{c}_{\mu}} \right)^{\nu} -
\frac{\gamma^{eff}_{\mu}}{2 N} \left(1-
\frac{g_{\mu}}{\bar{g}^{c}_{\mu}} \right)^{2\nu} \right]
\label{C18}
\end{equation}

Finally, using (\ref{C17}) and (\ref{C18}), we obtain the result for
$(\rho_{\parallel, \mu} - \rho_{\perp, \mu})/\rho_{\perp, \mu}$
to order $1/N$.
Reexpressing $\gamma_{\mu}$ in terms of correct
renormalization group invariants $\gamma_1$ and $\gamma_2$ of
(\ref{defgamma1}),
and exponentiating logarithmic terms,
we obtain the $1/N$ results for the crossover
exponents $\phi_1$ and $\phi_2$ that were given in (\ref{defphi1}).
Our value of $\phi_2$ coincides with the result by Lang and
Ruhl~\cite{Lang_Ruhl} who computed anomalous dimensions of
tensor fields of arbitrary rank
for critical $O(2N)$ sigma models. On the other hand, there do not
seem to be any other computations of $\phi_1$.

\subsection{Uniform susceptibility}
\label{unifchi}

The calculation of the uniform susceptibility
at small but finite $T$ and arbitrary
$\rho_{\perp}$ is essentially the same as that of $SU(N)$ response
at $T \rightarrow 0$; only the summation over frequency
should not be substituted
by the integration.  Doing the same calculations as have let us to
(\ref{C15}) but at finite $T$, we obtain to first order in $1/N$
\begin{equation}
\chi_{u} (T) = \left (\frac{g \mu_{B}}{\hbar}\right)^{2}
 ~\bar{\chi} (T) \left (1 +
\frac{\gamma_{\tau}}{2N}~ \bar{\chi} (T) \right)
\label{C20}
\end{equation}
Here
\begin{equation}
\bar{\chi} (T) = \bar{\chi} (T=0) + \delta \bar{\chi} (T)
\label{C21}
\end{equation}
where
\begin{equation}
\bar{\chi} (T=0) =  \frac{N}{2g_{\tau}}~
 \left( 1 + \frac{\gamma_{\tau}}{2N} \right)~
\left (1 - \frac{g_{\tau}}{\bar{g}^{c}_{\tau}}\right )^{\nu}
\label{C22}
\end{equation}
We expect that higher-order
corrections to (\ref{C20}) will only change $\gamma_{\tau}$ to
$\gamma^{eff}_{\tau}$.
 Notice that at $T \rightarrow 0$,
we recover a result consistent with (\ref{key4}) and (\ref{C9}):
\begin{equation}
\chi_{u} (T \rightarrow 0) = \left (\frac{g \mu_{B}}{\hbar}\right)^{2}
{}~\frac{2}{N}~\frac{N \chi_{\perp} + \chi_{\parallel}}{(N+1)}
\label{C22'}
\end{equation}
The temperature dependent piece $\delta \bar{\chi} (T)$
in (\ref{C20}) is precisely $1/2$
of that  in the isotropic $O(2 N)$ sigma model with $N-$dependent
spin-wave velocity $c^{*}$.
At $N= \infty$ we have from~\cite{CSY}
$\delta \bar{\chi} (T) = (k_{B} T /2 \pi c^{2}_{\perp})~f(x_1)$,
where numerically $ f(x_1)$ is close to 1
for all $k_{B} T/\rho_{\perp}$.
We will describe the structure of $1/N$ corrections
to $\delta \bar{\chi}$ and the value of $c^{\ast}$ in the following sections:
the $1/N$
results are of a rather different physical form depending upon whether
$k_{B} T \gg \rho_{\perp}$ or
$k_{B} T \ll \rho_{\perp}$. We will therefore consider the
 expressions for $\chi_u (T)$ and other observables separately in the
renormalized-classical and quantum-critical regions.

We now begin our discussion of various low-$T$ regions.
\section{renormalized-classical region}
\label{ren_cl}

This section will present expressions for different scaling functions in
the renormalized-classical region, $k_{B} T \ll
\rho_{\perp}$.   Under this condition, the low-temperature behavior
is related to the low-energy fluctuations of the macroscopic order parameter
 of the ground state and is
therefore  essentially classical. Indeed, this is true only for
fluctuations at sufficiently large scales
 when typical energies  $\hbar \omega \sim
\hbar c_{\perp}k \ll k_{B} T$,  and one need consider
only the $\omega_{n} =0$ term in the summation over Matsubara frequencies. At
larger $k$, quantum fluctuations are important, and  at $k >
\xi^{-1}_{J}$, antiferromagnet possesses $D = 2+1$ critical spin fluctuations
We  will consider this critical
behavior in the next section, and here  focus
 on the situation at small $\hbar c_{\perp} k < k_B T$. As in
unfrustrated antiferromagnets, there are two different low-$T$
regimes  already
in the classical region, because the actual (thermal) correlation
length $\xi$ is exponentially large when $k_{B} T \ll \rho_{\perp}$, and
one can have either
$k \xi \ll 1$ or $k \xi \gg 1$~~\cite{CHN}. Physically, the
crossover at $k \xi \sim 1$ is between the regime where the ordering is
destroyed by classical fluctuations and the dynamics is purely
relaxational ($k \xi <1$), and the regime where classical fluctuations are
 weakly damped propagating gapless spin-waves  ($k \xi
>1$). Below we will see how the spin structure factor
 changes in passing from one regime to the other. But first
we consider the behavior of the correlation length.

\subsection{Correlation length}
\label{xiren_cl}
As in the collinear case, we define the correlation length from
the equal-time, long-distance, $\exp(-r/\xi)$ decay of the
spin-spin correlation function.  From our previous discussion, especially
from (\ref{II5}) and (\ref{II6}), it is clear that the Fourier transform of
the spin correlator is related to the polarization operator rather than to
the Green function of the $z-$ field. At $N= \infty$, spinons behave as free
particles, and their propagator is $G_0 (k, i\omega) = 1/(k^2 + \omega^{2} +
m^{2}_0)$, where $m_0$ is the mass of the $z-$field, which at $N = \infty$
coincides with the inverse correlation length of the $O(2N)$ model. We then
obtain \begin{equation}
G ({\bf r}) \propto \int \frac{e^{i {\bf k r}} d^2 k ~ d^2 q}{\left[({\bf q} +
{\bf k}/2)^2 + m^{2}_0 \right]~\left[(({\bf q} -
{\bf k}/2)^2 + m^{2}_0 \right]} ~~ \propto e^{-2r m_0}
\end{equation}
  We see that in this limit,
the actual correlation length, $\xi$, is precisely $1/2 m_0$. We now proceed to
finite $N$. To first order in $1/N$, we have to consider
 self-energy and vertex corrections within a polarization bubble. A
 simple inspection of $1/N$ terms
shows that while self-energy
corrections renormalize the spinor Green function, and hence $\xi$,
  vertex corrections
  only modify the overall factor in the correlation function and do not
affect the exponent in the decay rate. In other words,
to first order in $1/N$, the actual correlation length is still precisely
a half  of that for the $z-$fields, and we therefore only have
to locate the pole in the zero-frequency part of the
 $z-$field propagator. For the isotropic case ($\gamma_{\mu} =0$), such
calculations have  already been performed in~\cite{CSY}. Here we have to
consider also  the effect of the $\gamma_{\mu}$ anisotropy. It is not difficult
to check
that the anisotropic term contributes to the self-energy to first order in
$1/N$, and therefore affects at this order the constraint equation
 which in essence is the equation  for $\xi$.
 The $\gamma$-dependent self-energy piece can easily be calculated
because the only nonvanishing contribution to order $1/N$ comes from the
diagram in Fig.\ref{figcurrent}. We obtain
\begin{equation}
\Sigma_{\gamma} (k, i\omega) = \frac{\gamma_{\tau}}{2N}~\omega^2 +
\frac{\gamma_{x}}{2N}~c^{2}_0 k^2
\label{C25}
\end{equation}
where $c_0 = c^{0}_{\perp} = \sqrt{\rho^{0}_{\perp} /\chi^{0}_{\perp}}$
Let us first keep  only anisotropic self-energy term. Substituting (\ref{C25})
 into the constraint equation (\ref{zmag})
and performing the momentum and frequency summation, we obtain
\begin{equation}
\frac{k_{B}T}{2 \pi} \log{\frac{k_{B}T}{\hbar c m}} =
{}~\frac{1}{g_x}~\left (1 -
\frac{g_x}{\bar{g}_c}\right)~\left (1 + \frac{\gamma_{x}}{2N}\right),
\label{C26}
\end{equation}
where $m$ is the full mass for the $z-$ field, $c$ is the linear combination of
the two spin-wave velocities which we will compute below,
and $\bar{g}_c$ is the same as in
(\ref{C15}).
We now observe that the r.h.s. of (\ref{C26}) can in fact be
reexpressed in terms of the fully renormalized transverse and longitudinal
spin-stiffness. Using (\ref{C15}) and (\ref{C18}), we find
\begin{equation}
\frac{k_{B}T}{2 \pi} \log{\frac{k_{B} T}{\hbar c m}} =
\frac{2 \rho_{\perp}}{N}~
\left (1 + \frac{1}{2(N+1)}~ \frac{(\rho_{\parallel} -
\rho_{\perp})}{\rho_{\perp}}\right)
\label{C27}
\end{equation}

Our next step is to determine how (\ref{C27}) is modified by
other $1/N$ corrections. We first consider the change
in the r.h.s. of (\ref{C27})
as $T \rightarrow 0$. At $\gamma_{\mu} =0$, earlier calculations
 to order $1/N^2$ ~\cite{CSY} have shown that the only temperature-independent
modification of the constraint equation
is the renormalization of the coupling  constant $ g_x$ in (\ref{C26})
to $g_{x}~(N-1)/N$. This renormalization can effectively be
regarded as the
wavefunction renormalization of the $z$-field, such that each $z-$ field
 propagator acquires a
factor $Z = (N-1)/N$.
 Physically, this renormalization is related to the fact that
the solution of the constraint equation at arbitrary small $T$ and finite
$\rho$ exists only for $N >1$, while for $N=1$ (i.e., for the XY case), a
single gapless spin-wave mode has no partner to interact with.
Consider now the $\gamma$-dependent piece in the r.h.s. of
(\ref{C27}). Clearly, it should also acquire an extra factor similar to
the renormalization og $g_{x}$. It is  difficult however
to determine explicitly
the $1/N$ renormalization of the anisotropic term because
 the anisotropic vertex itself has a
factor of $1/N$. On the other hand, the form of the
wavefunction renormalization seems quite plausible from a physical perspective,
 and we assume below, without
proof, that it remains the same in the anisotropic case as well.
  Simple considerations
then show that $\gamma_{x}$ should be substituted by $\gamma_x /Z = \gamma_x
N/(N-1)$.
We then obtain, keeping
only temperature-independent corrections in the r.h.s. of (\ref{C27})
\begin{equation}
\frac{k_{B}T}{2 \pi} \log{\frac{k_{B} T}{\hbar c m}} =
\frac{2 \rho_s}{N -1}~
\label{C28}
\end{equation}
where $c^2 = \rho_s /\chi$, and $\rho_s$ and $\chi$ are given by
\begin{eqnarray}
 \rho_{s} &=& \rho_{\perp} \left (1 + \frac{N}{2(N^2 -1)}~
 \frac{(\rho_{\parallel} -
\rho_{\perp})}{\rho_{\perp}}\right) \nonumber   \\
\chi &=&  \chi_{\perp} \left (1 + \frac{N}{2(N^2 -1)}~ \frac{(\chi_{\parallel}
-
\chi_{\perp})}{\chi_{\perp}}\right)
\label{C29}
\end{eqnarray}
Finally, we collect all temperature-dependent $1/N$ corrections to (\ref{C27})
using the same procedure as for the $O(2N)$ model~\cite{CSY}.
 These corrections include double
logarithms in the form $k_B T ~\log{\log{k_{B}T/m}}$, and regular $O(k_{B}
T)$ terms. Double-logarithms eventually give rise to the
temperature-dependent prefactor in $\xi$. Assembling all contributions,
 we finally obtain for the actual correlation length in
 frustrated antiferromagnet
\begin{equation}
\xi = \frac{1}{2}~\bar{\xi} \frac{\hbar c}{k_{B} T}~ \left(\frac{(N-1)
k_{B} T}{4 \pi \rho_s} \right)^{1/2(N-1)}~\exp\left[\frac{4 \pi \rho_s}{(N-1)
k_{B} T}\right]
\label{C30}
\end{equation}
where~\cite{Hasen}
\begin{equation}
\bar{\xi} = \left(\frac{e}{8}\right)^{1/2(N-1)}~ \times \Gamma (1 + 1/2(N-1))
\label{C31}
\end{equation}
We see therefore that, to first order in $\gamma$,
the expression for the correlation
length is the same, up to a factor of $1/2$,
 as in the $O(2N)$
 isotropic sigma-model with effective spin-stiffness
$4 \rho_s$ and spin-wave
velocity $c$. The factor of $4$ in $\rho_s$
 merely reflects the difference
between the definitions of the coupling constant $g$ in (\ref{C1}) and
in the $O(2N)$ $\sigma-$ model. At the same time, the overall factor
of $1/2$ is a signature of deconfined
spinons.
 For the physical case of $N=2$ we have
\begin{equation}
\rho_s = \frac{2}{3} \rho_{\perp} + \frac{1}{3} \rho_{\parallel}~~~
\chi = \frac{2}{3} \chi_{\perp} + \frac{1}{3} \chi_{\parallel}
\label{C32}
\end{equation}
To first order in $\gamma$ we also have $c = 2 c_{\perp}/3 +
c_{\parallel}/3$. The $T$ dependence in (\ref{C30}) then agrees
 with the two-loop renormalization group calculation of $\xi$
 performed by Azaria et al~\cite{Azaria}.
They also obtained the two-loop expression for the correlation
length in a classical model, valid at arbitrary ratio of the two bare
stiffnesses,
and argued that the result for the quantum case at arbitrary $\gamma_{\mu}$
will be the same if expressed
in terms of the fully renormalized $\rho_{\perp}$ and $\rho_{\parallel}$.
 Our analysis shows that this
 universal behavior of the correlation
length certainly exists to first order in $\gamma_{\mu}$, but we have no proof
that the universality persists at arbitrary $\gamma_{\mu}$.
 In any event, the analysis
presented here is valid close to the critical point when higher order
corrections due to anisotropy are small.

\subsection{Uniform susceptibility}
\label{chiren_cl}

The result for $\chi_u (T)$ valid at arbitrary ratio of $T/\rho_{\perp}$
is given by (\ref{C20}). We now use the
results of Ref~\cite{CSY} and obtain
\begin{equation}
\delta \bar{\chi} (T) =
\frac{N-1}{N}~ \frac{k_{B} T}{2 \pi c^{2}}
\label{C23}
\end{equation}
where $c$ is given by
(\ref{C29}).
 We expect that this result will hold at arbitrary $N$.
For the physical case of $N=2$ we then obtain using (\ref{C23})
and (\ref{C20}-\ref{C22})
\begin{equation}
\chi_{u} = \left(\frac{g \mu_{B}}{\hbar} \right)^2 ~\left(
\frac{\chi}{\chi_{\perp}} \right)~
\left[\chi_{\perp} + \frac{k_{B} T}{4\pi c^{2}} \right]
\label{C24}
\end{equation}
In Sec~\ref{applic} we will apply  our result  for $\chi_{u} (T)$ to the
$S=1/2$ Heisenberg antiferromagnet on a triangular lattice.

\subsection{Staggered susceptibility and structure factor}
\label{staggchiren_cl}

The key input for this subsection is our observation, in Eqns. (\ref{II3}),
(\ref{II6}),
 that the hydrodynamic
order-parameter variable for frustrated antiferromagnets is a
composite operator made of two $z-$fields, and
spin-spin correlation function is related to the polarization operator of
spinons.  At $N=\infty$, we use (\ref{II5}) and
 the results of Appendix \ref{appendt=0}, and
express $\chi_{s} (k , i\omega)$ as
\begin{equation}
\chi_s (k, i\omega) =  ~\frac{N^{2}_{0}}{4 \rho^{2}_{\perp}}~
 \Pi(k,i\omega)
\label{C33}
\end{equation}
Here $N_0$ and $\rho_{\perp}$ are the fully renormalized values
 of the on-site magnetization and spin-stiffness at $T=0$, and
$\Pi$ is the polarization  operator which for $\hbar c k , ~\hbar \omega \leq
k_B T$ is given by~\cite{CSY}
\begin{equation}
\Pi(k,i\omega) = \frac{k_{B} T}{\pi}~
\frac{\log{\left[\left(k^2 + \tilde{\omega}^2 +
\sqrt{(k^2 + \tilde{\omega}^2)^2 + 4 k^2 m^{2}_0}\right)/2 k m_0
\right]}}{\sqrt{(k^2 + \tilde{\omega}^2)^2 + 4 k^2 m^{2}_0}}
\label{C34}
\end{equation}
As before,
 $m_0$ is the mass of the $z-$field at $N= \infty$ which to this order
is also the inverse correlation length for the $z-$field, and $\tilde{\omega} =
\omega/c_0$.
We see that at small $k \leq m_0$, $\Pi (k, i\omega) \sim k_{B} T/4 \pi
m^{2}_0$ and hence $\chi_s (k,i\omega) \propto T \xi^2$,
 where $\xi$ is the
actual correlation length.
At the same time, at $k\xi \gg 1$, the logarithm in
the numerator of (\ref{C34}) cancels the overall factor of $T$, and we obtain
 $\chi_s (k,i\omega) \propto 1/k^2$ as it should be in the Goldstone regime.

We now consider how this simple behavior is modified by
$1/N$ corrections.  A simple inspection of the $1/N$ terms
shows that the diagrams which contribute to the renormalization of $\chi_s$
are  the same as in Sec~\ref{xiren_cl} - they include isotropic
 self-energy and vertex corrections
within a polarization bubble, and also corrections due to the $\gamma_{\mu}$.
 Let us first consider the isotropic case.
The self-energy corrections to the $z-$field at $\gamma_{\mu} =0$
 were studied in~\cite{CSY}. They give rise to
a renormalization of the mass and of the bare stiffness,
and also  yield an overall thermal
renormalization factor $\lambda_k$ for each Green function.
For $k \sim \omega/c_{\perp} \gg m$, this renormalization factor is
\begin{equation}
\lambda_k = \left(\frac{N-1}{N}\right)^{1/2}~
\left[\frac{\log[k_{B}T/(\hbar c_{\perp} m)]}{\log[\sqrt{k^2 +
m^2}/m]} \right]^{- 1/2(N-1)}
\label{C36}
\end{equation}
At $k \sim m$, the logarithm in the denominator is a number of the order of
one, and we have $\lambda_k =
[k_B T (N-1)/4 \pi \rho_{\perp}]^{1/2(N-1)}~ (1 +
{\cal O}(1/N))$. Further, it is not difficult to
check that the vertex renormalization
within a bubble also gives rise to logarithmic terms. Evaluating
the corresponding diagram in Fig.~\ref{figcorrpi}, to accuracy ${\cal O}(1)$
 and exponentiating the result,
 we obtain
 another renormalization factor $\zeta_k$, which at $k \gg m$ and to
order $1/N$ is
simply  $\zeta_k = \lambda^{2}_k$.
Collecting both contributions, we then obtain
\begin{equation}
\chi_s (k, i\omega) =  ~\frac{N^{2}_{0}}{4 \rho^{2}_{\perp}}~
\lambda^{4}_k ~\Pi(k,i\omega)
\label{C37}
\end{equation}
Finally, we consider the effect of the anisotropic term to first order
in $1/N$. Clearly, there are
 self-energy corrections to the $z-$ field propagators
 which eventually change $\rho_{\perp}$ to
$\rho_s$ and $c_{\perp}$ to $c$. Besides, the anisotropic terms
give rise to vertex corrections. We didn't perform  actual
calculations of the vertex corrections, but on general grounds
 it is likely that, to order ${\cal O}(1/N)$, they can be absorbed into the
renormalization of
 $N_0$. We then assemble (\ref{C28}),
(\ref{C34}), (\ref{C36}) and (\ref{C37}) and obtain
\begin{equation}
\chi_s (k, i\omega) = \frac{N^{2}_0}{\rho_s (N-1)}
 \left[\frac{k_B T (N-1)}{4 \pi \rho_s}\right]^{(N+1)/(N-1)}
{}~\xi^2 ~ f(k \xi, \omega \xi/c)
\label{C38}
\end{equation}
where the overall factor is chosen such that $f(0,0) =1$.
 It follows from
(\ref{C38}) that at finite $N$, $\chi_s (0,0) \propto \xi^2 ~
 T^{(N+1)/(N-1)}$.
This result is likely to be valid at arbitrary $N$.
For the physical case of $N=2$, it
reduces to $\chi^{aa}_s (0,0) \propto T^{3}\xi^2 $ - this is
 substantially smaller than
the naive mean-field result  $\chi_s (0,0) \propto T \xi^2 $.

 The behavior of $f(x,y)$ at intermediate $x,y =O(1)$ is rather
complicated, chiefly
because the spin-wave velocity also acquires a substantial downturn
renormalization at $k\xi =O(1)$~~\cite{CHN}. However
 at $k \xi \sim \omega \xi/c \gg 1$, the velocity renormalization is
irrelevant and using  (\ref{C36}) and (\ref{C37}) we obtain
\begin{equation}
f(x,y) = \left(\frac{N-1}{N+1}\right) ~\frac{1}{x^2 + y^2}~~
(\frac{1}{2}~\log(x^2 + y^2))^{(N+1)/(N-1)}
\label{C39}
\end{equation}
Substituting this result into  (\ref{C38}), and using the fact that at $k \xi
\gg 1, ~\log x \approx \log \xi$, we obtain to first order in $1/N$
\begin{equation}
\chi_s (k, i\omega) = \left(\frac{2}{N+1}\right)~\frac{N^{2}_0}{2\rho_s}~
\frac{1}{ k^2 + \omega^2 /c^{2}}
\label{C40}
\end{equation}
We now demonstrate that at any $N$, this expression is nothing but the
rotationally-averaged spin-wave result  for
the ordered  $SU(N)\times U(1)$ antiferromagnet.
 Indeed, using (\ref{II5}), (\ref{II6}) and (\ref{Zord}) we find that
the total number of transverse spin waves in the ordered state is
$N_{sw} = 2N$.  That (\ref{C40}) is
the averaged spin-wave result now follows from (\ref{II5}) and the fact
that each transverse spin-wave
mode at $T=0$ contributes a spin-wave factor
$N^{2}_0/2 \rho_s (k^2 + \omega^2 /c^{2})$ to $\chi_s$
(see Appendix \ref{appendt=0}).
 For the physical case of
 $N=2$, the averaging factor is $N_{sw}/N(N+1) = 2/3$, as it should be.

For experimental comparisons, it is also useful to have an expression
for the dynamical structure factor  $S(k, \omega)$ defined in (\ref{I7}),
and static
structure factor
\begin{equation}
S(k) = \int \frac{d \omega}{2 \pi}~ S(k, \omega)
\label{C42}
\end{equation}

As before, we will be interested in the behavior of $S (k, \omega)$ at
scales much larger than the Josephson correlation length. At such $k$,
quantum fluctuations are irrelevant, and using (\ref{I8}) and (\ref{I30}), we
can conveniently reexpress $S (k, \omega)$ as
\begin{equation}
S(k, \omega) = \frac{N^{2}_{0}}{2 \rho_s}~ \frac{ k_B T}{\rho_s}
{}~\frac{2 \hbar}{1
- e^{-\hbar \omega/(k_{B}T)}} \bar{\Xi}_1 (k, \omega)
\label{C43}
\end{equation}
where $\bar{\Xi}_1$ is related in a straightforward manner
to the universal function $\Xi_1$ introduced earlier in (\ref{I30}).
Below, we will restrict consideration of $ \bar{\Xi} (k, \omega)$
 to the frequency range $\omega \sim c_/\xi$, which is relevant
 for experimental comparisons in the renormalized classical region.

Consider first, the asymptotic behavior of $ \bar{\Xi} (k, \omega)$ at large
momentum $ k\xi \gg 1$. For such $k$, we found above that
 $1/N$ corrections to the polarization operator are
not singular. For a qualitative analysis, we can then
safely restrict ourselves
to $N=\infty$, when the imaginary part of the
polarization operator is given by~\cite{CSY}
\begin{eqnarray}
\mbox{Im} \Pi (k, \omega) &=& \frac{(\hbar c)^4}{4 \pi}~
\int~\frac{d^2 q}{4 \epsilon_{q_1}
 \epsilon_{q_2}}~ [(1 + n_{q_1} + n_{q_2})
{}~(\delta(\epsilon_{q_1} + \epsilon_{q_2} - \hbar \omega) -
{}~(\delta(\epsilon_{q_1} + \epsilon_{q_2} + \hbar \omega)) + \nonumber \\
&& (n_{q_1} - n_{q_2})~(\delta (\epsilon_{q_2} -
\epsilon_{q_1} -\hbar \omega) - \delta (\epsilon_{q_2} -
\epsilon_{q_1} +\hbar \omega))]
\label{C44}
\end{eqnarray}
Here $n_q$ is a Bose function and  $\epsilon_q = (\hbar c) \sqrt{q^2 +
\xi^{-2}}$.  At $c k \gg \omega$, the only contribution to
$\mbox{Im} ~\Pi (k, \omega)$ comes from the second piece in (\ref{C44}),
which describes collisionless Landau damping. Doing the integration,
 we obtain a simple result
\begin{equation}
\bar{\Xi}_1 (k, \omega) = \frac{\omega}{2 \pi c k^3}.
\label{C45}
\end{equation}
Note that as expected, $\bar{\Xi}_1 (k ,\omega)$ scales linearly with
$\omega$.

We turn now to a discussion of smaller $k$.
The corrections to (\ref{C44})
include the terms similar to $\lambda_k, \zeta_k$ above, which
grow logarithmically with decreasing $k$ and eventually change the temperature
dependence of $\bar{\Xi}_1$ at $ k\xi \sim 1$.
 Moreover, at such momenta, the damping
of excitations becomes comparable to the real part of the quasiparticle energy,
and we cannot simply restrict ourselves to collisionless
Landau damping.
We did not perform explicit $1/N$
calculations of $\bar{\Xi}_1$ at intermediate $k$, but for an estimate we
can rely on the results of Ref~\cite{Tyc_Halp,CSY}
for the momentum dependence of the damping of excitations in the
$O(2N)$ sigma-model.
Combining these results with the expressions (\ref{C38}, \ref{C39})
 for the real part of $\chi_s$, we obtain
\begin{equation}
 \bar{\Xi}_1 (k, \omega) \propto \frac{\hbar \omega ~\gamma_{k \omega}}{ \hbar
 (c_{k} k^2)^2}
\left(\frac{\rho^{k}_s}{\rho_s}\right)^{2/(N-1)}~
\label{C46}
\end{equation}
Here $\gamma_{k \omega}$ is the damping of $z$-field excitations  given
by~\cite{Tyc_Halp,CSY}~
 $\gamma_{k \omega} \propto \hbar c_{k} k ~(k_{B} T/\rho^{k}_s)^2
{}~\log{\rho^{k}_s /k_B T}$, and the
momentum-dependent spin-stiffness and spin-wave velocity are introduced
as another way to account for the
 logarithmical terms in (\ref{C39}):
\begin{equation}
\rho^{k}_s = \frac{(N-1) k_{B} T}{4 \pi} \log {k \xi};~~~~~
(c_{k})^2 \propto   \rho^{k}_s
\label{C47}
\end{equation}
(We note in passing that  at $k \xi \gg 1$, we have with the logarithmical
accuracy $ \rho^{k}_s =
\rho_s, c_{k} = c$.)  At $k \xi = {\cal O}(1)$, we have
$\rho^{k}_s \propto T$, $c_{k} \propto \sqrt{T}$ at {\it arbitrary} $N$,
 and hence, our final result
\begin{equation}
\bar{\Xi}_1 (k, \omega) \propto  \frac{\omega \xi^3}{c}~
\left(\frac{(N-1) k_B
T}{4 \pi \rho_s}\right)^{(5-N)/2 (N-1)}~
\label{C48}
\end{equation}
For $N=2$, we have $\bar{\Xi}_1 (k, \omega) \propto \omega ~T^{3/2}$.

Finally, we consider the static  structure factor, $S(k)$. A simple
inspection shows that the frequency integral in (\ref{C42}) has
 two basic contributions. One comes from large $\omega$ where the system is
$D=2+1$ critical, while the second comes from $\hbar \omega < k_B T$
where the fluctuations are essentially classical.
We will see in the next section that at large $\omega$, $\Pi (k, \omega)$
behaves as $1/ \omega^{2-{\bar \eta}}$ where ${\bar \eta}$ is given by
(\ref{exponents}). We found earlier
 that ${\bar \eta} >1$ (at least, at large $N$,
and hence the frequency integral over quantum fluctuations
explicitly depends on the upper cutoff in the theory.
We will discuss nonuniversality in $S(k)$ in more detail in the next section.
In the renormalized-classical region however, the correlation length is
exponentially large and we may expect that the dominant contribution to
$S(k)$, which scales as $\xi^2$, still comes from the frequences
$\omega \propto \xi^{-1}$ where fluctuations are essentially classical.
 For such frequences, the rescaling factor between $\mbox {Im}~\chi (k,
\omega)$ and $S (k, \omega)$
is $2/(1 - e^{\hbar \omega /k_{B} T}) \approx 2k_B T/\hbar \omega$,
 and we have simply $S(k) = k_B T \chi_s (k,0)$, where $\chi_s (k, 0)$
is given by (\ref{C38}). At $k=0$ we then obtain $S(0) \propto T^{2
N/(N-1)}~\xi^2$. For $N=2$, this reduces to $S(0) \propto T^4 \xi^2$.

\subsection{Local susceptibility and spin-lattice relaxation rate}
\label{t1ren_cl}
Another experimentally measured quantity is the momentum-integrated
dynamical structure factor $S(\omega) = \int d^2 k ~S(k, \omega) /4 \pi^2$.
Unlike $S(k)$, this observable is universal in 2D as can easily be seen from
(\ref{C45}). It is also not difficult to show that for $\omega \sim
c \xi^{-1}$, the integration over momentum is
confined to $k \sim \xi^{-1}$, where we can use
the estimate (\ref{C43}, \ref{C48}) for $S(k, \omega)$. We then obtain
\begin{equation}
S_L (\omega) \propto \frac{N^{2}_{0} \xi}{c}~ \left(\frac{(N-1) k_B
T}{4 \pi
\rho_s}\right)^{(3N +1)/2(N-1)}
\label{C49}
\end{equation}

Further, the $\omega \rightarrow 0$ limits of $S(k, \omega)$ and
$\chi_ (k, \omega)$ are related to the transverse ($1/T_1$)  and
longitudinal ($1/T_2$) relaxation rates for
nuclear spins coupled to electronic spins in the antiferromagnet.
We have
\begin{eqnarray}
\frac{1}{T_1} &=& 2~\lim_{\omega \rightarrow 0}~
\int \frac{d^2 k}{4 \pi^2 \hbar^2} A^{2}_k  ~S(k, \omega)\\
\left(\frac{1}{T_2}\right)^{2}_{NMR} &=& 2~\lim_{\omega \rightarrow 0}~
\left(\frac{\rho_s}{\hbar c}\right)^2 ~\int
\frac{d^2 k}{4 \pi^2 \hbar^2} \bar{A}^{4}_k  ~\chi^{2}_s (k, \omega)
\label{C50}
\end{eqnarray}
where $A_k$ and $\bar{A}_k$ are the hyperfine coupling constants
(with the dimension
 of energy). They generally tend to some finite values as $k \rightarrow 0$.
The factors of $2$ appear because fluctuation modes near $Q$ and $-Q$
equally contribute to relaxation rates.
The temperature dependence of $1/T_1$ then immediately
follows from the result (\ref{C49}) for $S(\omega)$. For $N=2$ we obtain
\begin{equation}
\frac{1}{T_1} \propto
\left(\frac{A_0}{\hbar}\right)^2 ~\frac{N^{2}_0 \xi}{c}~\left(\frac{k_B
T}{\rho_s}\right)^{7/2}
\label{C51}
\end{equation}
An exactly parallel computation can be done for the spin-echo decay
rate $1/T_2$,  and the result is (for general $N$)
\begin{equation}
\frac{1}{T_2} \propto \left(\frac{\bar{A}_0}{\hbar}\right)^2
{}~ \frac{N^{2}_0 \xi}{\hbar^2 c}~\left(\frac{k_B
T}{\rho_s}\right)^{(N+1)/(N-1)}
\label{C52}
\end{equation}
For $N=2$ this yields $T^{-1}_2 \propto T^{3} \xi$.

\section{quantum-critical region}
\label{qc}

We now consider the results for the quantum-critical region where $4 \pi
\rho_s < N k_{B} T$. Under this condition, the relevant scale for
fluctuations is given by $T$ itself and both quantum and classical
fluctuations are equally important (i.e., at relevant energies, Bose functions
are {\cal O}(1)). Our first observation
in this region concerns the role of the anisotropic ({\em i.e.\/}
$\gamma_{\mu}$
dependent) terms in the action. The scaling hypothesis predicts that any
scaling
function near the quantum transition should depend on the dimensionless ratio
 $\xi_{J}/L_{\tau}$ where $\xi_J$ is the Josephson correlation length, and
$L_{\tau} = \hbar c/k_{B} T$ is a finite length in the imaginary
time direction at $g = g_{c}$. We
have shown above in Sec \ref{cons_curr} that at $T=0$, anisotropic corrections
 had a form $\gamma_{\mu} (\xi_J /a)^{-\phi_{1,2}}$ where both crossover
exponents are clearly positive and even
larger than $1$ at finite $N$ (see Eqn. (\ref{defphi1})).
We therefore expect that the leading
anisotropic corrections deep in the quantum-critical region will scale as
 $\gamma_{\mu} (k_B T a/\hbar c)^{\phi_{1,2}}$ with positive
$\phi_{1,2}$, i.e.  they will be
subdominant at low $T$ compared to the leading  terms in
the scaling functions.
Clearly then,
the quantum-critical behavior will  be the same as in the
isotropic $O(2N)$ sigma-model. The anisotropic term in the action
 will however  renormalize the
spin-wave stiffnesses
 and velocities in the subleading terms in the full scaling
functions, which describe deviations from the pure critical behavior.
These terms will be calculated in this Section only at $N= \infty$, at
which order the anisotropic term in the action does not contribute. We will
then
assume, without proof, that the renormalization due to $\gamma_{\mu}$ leads to
the same effective $\rho_s$ and $\chi$ given by (\ref{C29})
 as the renormalized-classical
expressions. On general grounds, this is likely to be the case because
the corrections to the pure quantum-critical formulas
 account for the crossover to the renormalized-classical region. However,
as we said, explicit calculation of the subleading terms at finite
$N$ has not been performed.

  We emphasize that even in the absence of the anisotropy,
 the scaling properties of spin correlators are  quite
different from those for unfrustrated antiferromagnets simply
because each spin component is a bilinear product of the
$z-$fields. We now consider separately the behavior of various
observables.

\subsection{Correlation length}
\label{xiqc}
The expression for the correlation length follows directly from the
observation that the spin propagator  is a convolution of  two Green functions
for $z$-fields. An analysis, similar to that for the renormalized-classical
region,
shows that vertex corrections in the polarization  bubble do not effect the
form of
the exponential decay of correlations, and therefore the actual correlation
length is
again exactly 1/2 of that for the $O(2N)$ sigma-model.  Specifically, we obtain
\begin{equation}
\xi (T) = \frac{1}{2}~ \frac{\hbar c}{k_{B} T}~ X_1 (\infty) \left[1 -
\kappa {\bar{x}_1}^{-1/\nu} + \ldots \right]
\label{Q3}
\end{equation}
where  we defined $\bar{x}_1 = N k_B T/4 \pi \rho_{s}$, and $\nu$ is the
exponent for the Josephson correlation length given by (\ref{exponents}).
The values of $X_1 (\infty)$ and $\kappa$ were
 found earlier~\cite{CSY}: $X_1 (\infty)
 = \Theta (1 + 0.1187/N)$, where
 $\Theta = 2 \log{(1 + \sqrt{5})/2} = 0.962424$, and $\kappa = 2/\sqrt{5} +
O(1/N)$.

\subsection{Uniform susceptibility}
\label{chiqc}
We continue with the response to the uniform magnetic field. As in the
renormalized-classical region, we
use the general result (\ref{C20}), but now the temperature dependent
piece in $\bar{\chi}$ is dominant, and to order $1/N$ the universal function
for $\chi_u$ defined in (\ref{I34}) is given by
\begin{equation}
\Omega (x_1, y_{\rho}, y_{\chi}) = \frac{\sqrt{5}}{2 \pi}~\log{\frac{\sqrt{5} +
1}{2}}~
 ~\left[\left(1 - \frac{0.31}{N}\right) +
\alpha {\bar{x}_1}^{-1/\nu} + \ldots \right]
\label{Q1}
\end{equation}
where $\alpha = 0.8 + O(1/N)$.
This is indeed a half of the susceptibility for  $O(2N)$ square-lattice
antiferromagnet.
At $N=2$, we obtain using the mean-field $(N=\infty)$ result for the correction
term
\begin{equation}
\chi_u (T) = \left(\frac{g \mu_B}{\hbar}\right)^2 ~ \left[0.86 \chi
 + 0.145 \frac{k_{B} T}{c^{2}}\right]
\label{Q2}
\end{equation}

\subsection{Dynamic susceptibility and structure factor}
\label{dynchiqc}

In this subsection, we compute the scaling functions $\Phi_{1s} (\overline{k},
\overline{\omega}, x={\infty}, y_{\rho} =0, y_{\chi} =0) =
\Phi_s (\overline{k},
\overline{\omega})$ and
$\Xi_1 (\overline{k}, \overline{\omega}, x ={\infty},
y_{\rho} =0, y_{\chi} =0) =
\Xi (\overline{k},
\overline{\omega})$
for staggered
dynamical susceptibility and structure factor at the critical point $g =g_c$.
These two scaling functions were introduced in  (\ref{I30}) and are related by
the fluctuation-dissipation
 theorem $\Xi (\overline{k}, \overline{\omega}) =
{}~\mbox{Im} \Phi_s (\overline{k}, \overline{\omega})$.
 As before,  $\Phi_{s} (\overline{k}, \overline{\omega})$ is
simply related to the polarization operator for $z-$ fields:
$\Phi_{s} (\overline{k}, \overline{\omega}) = (k_B T/2 (\hbar c_{\perp})^2)
{}~\Pi (k, \omega)$.
The limiting behavior of $\Pi$ and hence
$\Phi_s (\overline{k}, \overline{\omega})$ at small and large
$\overline{k}$ and $\overline{\omega}$ can be obtained by properly expanding
the real part of (\ref{C34}). For large
$\overline{k}, \overline{\omega}$, we found using the results
of~\cite{CSY,subir}
\begin{equation}
\Phi_s (\overline{k}, \overline{\omega}) =
\frac{1}{16 \sqrt{\overline{q}^2 -
(\overline{\omega} + i \delta)^2}} + \frac{8 \zeta (3)}{5 \pi}~
\frac{2 (\overline{\omega}^2 +
\overline{q}^2)}{(\overline{k}^2 - (\overline{\omega}+ i \delta)^2)^3} +
O\left(\frac{1}{(\overline{k},\overline{\omega})^6}\right)
\label{Q4}
\end{equation}
On the other hand, at small momentum and frequency, an expansion in
the real part of (\ref{C34}) yields
\begin{equation}
\mbox{Re}~
 \Phi_s (\overline{k}, \overline{\omega}) = \frac{\sqrt{5}}{16 \pi \Theta}
{}~\left( 1 - \frac{\overline{k}^2 (1 + 2 \Theta/\sqrt{5}) -
\overline{\omega}^2}{12 \Theta^2} +
O\left((\overline{k},\overline{\omega})^4\right) \right)
\label{Q5}
\end{equation}

We consider next $1/N$ corrections to these results.
At small $\overline{k}$ and $\overline{\omega}$, the expansion in
$1/N$ does not involve logarithms. Regular
$1/N$ corrections to $\Phi_s$ were found to be quite small for
unfrustrated  antiferromagnets~\cite{CSY} and we expect the same to
be true in our case as well. On the other hand, at
$\overline{k}, \overline{\omega} \gg 1$, the behavior is nearly the
same as at the critical point at $T=0$,
and using the results of Appendix \ref{appendt=0} we found that the
leading term in (\ref{Q4}) is modified to
\begin{equation}
\Phi_s (\overline{k}, \overline{\omega}) = \frac{A_{N}}{16
(\overline{q}^2 - (\omega + i \delta)^2)^{1 - {\bar \eta}/2}}
\label{Q6}
\end{equation}
where $A_{N} = 1 +O(1/N)$ and ${\bar \eta}$ is given by  (\ref{exponents})

It is also not difficult to compute explicitly the
imaginary part of the polarization operator, which
 then yields the scaling function for the dynamic structure factor.
In the two assymptotic limits of
large and small $\overline{\omega}$ we obtained
\begin{equation}
\Xi (\overline{k}, \overline{\omega}) =
\frac{A_N \sin(\pi{\bar \eta}/2)}{16}~\frac{\theta (\overline{\omega}^2 -
\overline{k}^2)}{(\overline{\omega}^2 - \overline{k}^2)^{1 - {\bar \eta}/2}}
\label{Q7}
\end{equation}
for  $\overline{\omega} \gg 1$  and
\begin{equation}
\Xi (\overline{k}, \overline{\omega}) =
\frac{\overline{\omega}}{8
\sqrt{\pi}}~\frac{\exp{-\overline{k}/2}}{\overline{k}^{3/2}}
\label{Q8}
\end{equation}
for $ \overline{\omega} \ll 1$ and $\overline{k} \gg 1$. In (\ref{Q7}), $\theta
(x)$ is a step function. It is also not
difficult to obtain the $N = \infty$ expression for $\Xi$ for both
$\overline{\omega} \ll 1$ and $\overline{k} \ll 1$, but in this region of
momentum and frequency, quasiparticle excitations are overdamped, and
one again
cannot restrict to the $N=\infty$ result of
 collisionless Landau damping. We can only expect
on general grounds that at small $\overline{\omega}$ and arbitrary
$\overline{k},~
 \Xi (\overline{k}, \overline{\omega}) \propto \overline{\omega}$.

We consider, further, the static structure factor $S(k)$ defined by
(\ref{C42}).
In the quantum-critical region, we have  at $ g \nearrow g_c$
\begin{equation}
S (\overline{k}) = \frac{N^{2}_0}{2}~\left(\frac{\hbar c}{\rho_s}\right)^2
{}~\left(\frac{N k_B T}{4 \pi \rho_s}\right)^{{\bar \eta} -1}
{}~ I(\overline{k})
\label{Q9}
\end{equation}
where $I(\overline{k}) =
 \int d \overline{\omega} ~(1 - e^{-\overline{\omega}})^{-1} ~
\Xi (\overline{k}, \overline{\omega})/ \pi$.
Notice that the functional form of $S(\overline{k})$ is similar to that for
unfrustrated antiferromagnets~\cite{CSY}. Moreover, in both frustrated and
unfrustrated cases,
$\Xi (\overline{k}, \overline{\omega}, \infty)$ behaves
 at large frequences as $\Xi (\overline{k}, \overline{\omega}, \infty) \propto
(\overline{\omega})^{-2 + {\bar \eta}}$. The difference between the two cases
is in the value of $\bar{\eta}$. For unfrustrated antiferromagnets,
${\bar \eta} \approx 0$ and the frequency integral in $I(\overline{k})$ is
convergent. For frustrated systems, ${\bar \eta} > 1$ at least, at large $N$
(see (\ref{exponents})),
 and the integral over frequency in $I (\overline{k})$
is divergent, which actually means that
 the dominant contribution to $S(\overline{k})$ at small
temperatures comes  from the frequencies of the order of a cutoff.
 Specifically, using (\ref{C34}) we obtain
\begin{equation}
I(\overline{k}) = \frac{B_N}{{\bar \eta}-1}
{}~\left[\left(\bar{\Lambda}^{{\bar \eta}-1}~ -
\Theta^{{\bar \eta}-1}\right)\right] + I^{\prime}
(\overline{k})
\label{Q10}
\end{equation}
Here $B_N = 1 + {\cal O}(1/N)$, $\Theta$ was defined after (\ref{Q3}),
$\bar{\Lambda} = \hbar c \Lambda /k_B T$ where
$\Lambda$ is a relativistic cutoff in the theory,
 and  $I^{\prime} (\overline{k})$ is a universal function of momentum, which
as $\overline{k} \rightarrow 0$ tends to $I^{\prime} (0) = 1.67 +O(1/N)$.
The nonuniversality in
$I(\overline{k})$ at low temperatures is now
transparent.
Recall however that (\ref{Q10}) is valid only
 in  the quantum-critical regime where the temperature is the only scale
for fluctuations. In the renormalized-classical regime, the analog of
$I^{\prime} (\overline{k})$ is proportional to the square of the actual
correlation length $\xi$, and  is exponentially large at low $T$
compared to the nonuniversal piece in $I (\overline{k})$, which does not
contain
any dependence on $\xi$.

\subsection{Local susceptibility and spin-lattice relaxation}
\label{locchiqc}

Unlike $S(\overline{k})$, the  momentum-integrated
dynamic structure factor
$S_L (\omega) = \int d^2 k ~S(k, \omega) /4 \pi^2$
is universal in $2d$, as we already found in the renormalized-classical region.
The scaling function for $S_L (\omega)$ was introduced in (\ref{I35}).
At $g = g_c$ and $N = \infty$,
 this scaling function  can be deduced
directly from (\ref{C34}). A simple calculation yields
\begin{equation}
K(\overline{\omega}) = \frac{1}{32 \pi}~
\left[2 \log{\frac{1 - e^{-(\Theta + \overline{\omega})}}{1 - e^{-\Theta}}}
+ \Theta (\overline{\omega} - 2 \Theta)~\left ( \overline{\omega} - 2 \Theta +
2  \log{\frac{1 - e^{-(\overline{\omega} - \Theta)}}{1 -
e^{-\Theta}}}\right)\right]
\label{Q10'}
\end{equation}
where $K (\overline{\omega}) = K_1 (\overline{\omega}, x_1 = \infty,
, y_{\rho} =0, y_{\chi} =0)$. At small $\overline{\omega}$, this reduces to
\begin{equation}
K (\overline{\omega}) = \frac{\sqrt{5} -1}{64\pi} ~\overline{\omega}
\label{Q11}
\end{equation}
while at $\overline{\omega} \gg 1$, we have
\begin{equation}
K (\overline{\omega}) = \frac{\overline{\omega}}{32 \pi}
\label{Q12}
\end{equation}
Note that linear dependence on $\overline{\omega}$ is present in both limits
(in the unfrustrated case, $K (\overline{\omega})$ saturated at large
$\overline{\omega}$).

We now consider $1/N$ corrections to these results. At $\overline{\omega} \ll
1$, the expansion in $1/N$ is free from divergences because each $z-$ field
propagator has a gap $\Theta \sim 1$. The expansion in $1/N$ then holds
in integer powers of $1/N$, and numerically we expect the corrections to
 (\ref{Q11}) to be quite small. On the
contrary, at large $\overline{\omega} \gg 1$, the actual
form of $K (\overline{\omega})$
is different from the $N=\infty$ result because of
the singular $1/N$ corrections. Using (\ref{Q7}) we obtain instead of
(\ref{Q12})
\begin{equation}
K (\overline{\omega}) = \frac{A_N \sin
(\pi{\bar \eta}/2)}{32 \pi} \frac{\overline{\omega}^{\bar \eta}}{\bar \eta}
\label{Q13}
\end{equation}

Finally, the $\overline{\omega} \rightarrow 0$ limit
of $S_L (\overline{\omega})$ is related to the
transverse spin-lattice relaxation rate. As before, we assume that the
hyperfine coupling constant $A_k$ tends to a finite value at $k =0$ and
the dominant contribution to $1/T_1$ thus comes from the momentum range
$\overline{k} = O(1)$. Using (\ref{I35}), (\ref{C50}) and (\ref{Q11}),
 we then obtain
\begin{equation}
\frac{1}{T_1} = \left(\frac{A_0}{\hbar}\right)^2 ~ Z~\frac{N^{2}_0 ~\hbar}
{\rho_s}~~\left(\frac{N k_B T}{4
\pi \rho_s}\right)^{\bar \eta}
\label{Q14}
\end{equation}
where $Z = (\sqrt{5} -1)/8 N  + O(1/N^2)$.

A parallel analysis can be done for longitudinal spin-lattice relaxation
$1/T_2$ defined in (\ref{C50}), and the result is
\begin{equation}
\frac{1}{T_2} \propto \left(\frac{\bar{A}_0}{\hbar}\right)^2~
 \frac{N^{2}_0 ~\hbar}
{\rho_s} ~\left(\frac{N k_B T}{4 \pi \rho_s}\right)^{{\bar \eta} -1}
\label{Q15}
\end{equation}

\section{quantum disordered region}
\label{qd}

Let us first notice some crucial properties
of the  quantum-disordered ($g > g_c$) phase at $T=0$. The
presence of free spin-1/2 $z$ quanta implies that $\chi_{s} ( k,
\omega)$ only has a branch cut in the complex $\omega$ plane. This
should be contrasted with the behavior of collinear
antiferromagnets~\cite{CSY} which had an additional spin-1
quasiparticle pole. We will compute below the structure of this
branch cut at $N=\infty$.

Before doing this, it is useful to introduce our precise definition
of the prefactor ${\cal A}$. We will use the $T=0$ form of the local
dynamic structure factor $S_{L} ( \omega )$ to specify the
normalization. Using the fact that the dynamic susceptibility
involves a response function with 2 spinon intermediate states, each
with a gap $\Delta$, and that each spinon propagator has a
quasiparticle pole we can show quite generally (to all orders in
$1/N$ or $\epsilon=4-D$) that near the threshold in the quantum
disordered phase we must have
\begin{equation}
S_{L} ( \omega) = {\cal A} \frac{\hbar \omega - 2 \Delta}{2 \Delta}
\theta(\hbar \omega - 2 \Delta)
{}~~~~~~\mbox{$\omega$ close to $2 \Delta$}.
\end{equation}
The above form {\em
defines\/} the values of ${\cal A}$ and $\Delta$.
Combined with $c$, these parameters universally determine the entire
staggered susceptibility.

The $N=\infty$ computation of $\chi_s$ is standard~\cite{CSY}. The $z$
propagators acquire a gap $\Delta$ given by
\begin{equation}
\Delta = 4 \pi \left( \frac{1}{g_c} - \frac{1}{g} \right)
\end{equation}
We evaluated the susceptibility using (\ref{II5}) and found at
$N=\infty$
\begin{equation}
\mbox{Im} \chi_s (k, \omega ) =
\mbox{sgn} ( \omega ) \frac{{\cal A} \pi c^2}{2
\Delta} \frac{1}{\sqrt{\omega^2 - c^2 k^2}} \theta\left( \omega^2  -
c^2 k^2 - 4 \Delta^2 / \hbar^2 \right)
\end{equation}
with ${\cal A} = g^2 \Delta / (4 \pi \hbar^2 )$. These
results are consistent with (\ref{deltanu}) and (\ref{extr}) provided
$\nu=1$ and $\bar \eta = 1$.
Note that $\chi_s ( k , \omega )$ has branch cuts emanating from
$\pm (4 \Delta^2 / \hbar^2 + c^2 k^2 )$ to $\pm \infty$.
Compare this result with the confined
spinon model of (\cite{CSY}) where
at $N=\infty$, $\mbox{Im} \chi_s$ was simply a delta function.

It is simple to extend the above results for $\chi_s$ to finite
temperature and to order $1/N$. The main effect of small $T$ is to
fill in the gap in the spectrum by exponentially small terms. The
$1/N$ corrections do not introduce any essentially new features, and
will not be considered here.

\section{application to a $S=1/2$ antiferromagnet}
\label{applic}

In this section,  we compare our scaling results
 to the properties of the $S=1/2$ Heisenberg antiferromagnet
on a triangular lattice. As input, we need the values of sublattice
magnetization, two spin stiffnesses
and two spin-susceptibilities at $T=0$. In Appendix \ref{largeS}, we have
calculated
these parameters in the $1/S$ expansion, to order $1/S$ for the stiffnesses and
susceptibilities and to order $1/S^2$ for sublattice magnetization.
 The extension of our large $S$
results to $S=1/2$ yields
\begin{equation}
N_0 = 0.266; ~\chi_{\perp} = \frac{0.091}{J a^2}~\chi =
\frac{0.084}{J a^2}, ~
\rho_s = 0.086J, ~ c = \sqrt{\rho_s /\chi} = 1.01J a
\label{S1}
\end{equation}
The magnitude of the $1/S^2$
result for the sublattice magnetization
indicates that higher-order corrections are rather small.

Let us now summarize the scaling predictions which follow from the values in
(\ref{S1}).
For the uniform susceptibility, we obtain
\begin{equation}
\chi_{u} = \left(\frac{g \mu_{B}}{\hbar a} \right)^2 ~\frac{1}{J}~\left[
0.084 + 0.07 \frac{k_{B} T}{J} \right]
\label{S2}
\end{equation}
in the renormalized-classical regime, and (using the $N=\infty$ result for the
correction term)
\begin{equation}
\chi_{u} = \left(\frac{g \mu_{B}}{\hbar a} \right)^2 ~\frac{1}{J}~\left[
0.072 + 0.14 \frac{k_{B} T}{J} \right]
\label{S3}
\end{equation}
in the quantum-critical regime.
Comparing (\ref{S2}) and (\ref{S3}), we observe
 that the slope of $\chi_u$ in the
quantum-critical regime is nearly twice as large as in (\ref{S2}), while
the value of the intercept is larger in the renormalized-classical regime.
Further, the correlation length behaves in the renormalized-classical regime
as
\begin{equation}
\xi \approx 0.24 \left (\frac{4 \pi \rho_s}{ k_B T}\right)^{1/2}~
\exp[4 \pi \rho_s /k_B T]
\label{S4}
\end{equation}
where $4 \pi \rho_s \approx 1.08 J$, and
deep in the quantum-critical region as
\begin{equation}
\xi = \frac{0.51 J a}{k_B T}
\label{S5}
\end{equation}
Finally, in the renormalized-classical regime, the universal contribution to
$S(k)$ is dominant, and for $k=0$ we obtain from (\ref{C38}) and (\ref{C42})
\begin{equation}
S(0) \approx 0.85 \left(\frac {k_B T}{4 \pi \rho_s}\right)^4 \xi^2
\label{S5'}
\end{equation}
In the quantum-critical region, the dominant piece in $S(0)$ is a
temperature-independent contribution from lattice scales, and we can only
conclude that deep inside quantum-critical region, $ S(0) \sim A +
 B (T/T_0)^{\bar{\eta} -1}$, where $A$ is a
$T$ -independent nonuniversal piece.
Using the large $N$ results we
found $B \approx -0.27 a^2$ and $T_0 \approx 0.54 J$.

{}From the discussion in the bulk of the paper, we expect the crossover between
the classical and quantum regimes to occur somewhere around $x_1 =1$ i.e. at
$ k_B T = 2 \pi \rho_s \sim 0.5J$. This indeed is not a very small crossover
temperature. However, the analysis for the unfrustrated case~\cite{CSY}
shows that the uniform susceptibility
displays quantum-critical behavior starting already below $x_1 =1$. We
therefore
first compare our results with the numerical data on uniform susceptibility.

The temperature dependence of $\chi_u$ was recently studied in
 high-temperature series expansions for $S=1/2$ triangular
antiferromagnet~\cite{Young}. The data show
that $\chi_u$ obeys a Curie-Weiss law at high
$T$, passes through a maximum at $T \approx 0.4J$, and then falls down.
In general, the temperature where $\chi_u$ has a maximum roughly separates
the low-temperature region below the maximum
where a long-wavelength approach is valid,
from the high-temperature
region where the physics is dominated by lattice-scale effects.
It is unfortunate that this
this temperature is rather low for the triangular antiferromagnet,
because it reduces substantially
the temperature range for low-energy behavior (for comparison, in the
square-lattice antiferromagnet, the maximum in $T_c$ occurs at $k_{B} T \sim
J$).
Numerical data~\cite{Young} is available only over a small $T$ region below the
maximum. Nevertheless, we fitted
 the data by
a linear in $T$ dependence and found $0.13 \pm 0.03$ for the slope and around
$0.06$  for the
intercept - both results are in reasonable agreement
with our quantum-critical expression (\ref{S3}).
Finally, at very low $T$, we expect a crossover to the renormalized-classical
regime, and the $T=0$ value in (\ref{S2}) is also consistent with the data.

Now about $S(0)$. Previous studies of square-lattice antiferromagnets
have shown that  we can hardly expect to observe
pure quantum-critical behavior for $S(0)$ at $x_1 \sim 1$.
Indeed, the leading correction to $S(0)$ due to the deviation from purely
quantum-critical behavior is  $\delta S(0) = C
(2 \pi \rho_s / k_B T)$, where at $N = \infty$ we found  $C = 0.45 a^2$.
Clearly then, at $ k_B T \sim 0.5 J$,
temperature dependence related to
deviations from pure criticality is likely to overshadow the weak
($(T/T_0)^{{\bar \eta} -1}$) temperature
dependence  in $S(0)$ at $g = g_c$; instead, we expect that at such
$T$, the structure factor should roughly follow
$S(0) = A + C (2 \pi \rho_s /k_B T)$, or (still considering second term as a
correction)  $k_B T \log {S(0)} \approx  2 \pi \rho_s C/A + k_B T \log{A}$.
The series expansions~\cite{Young} yielded $k_B T \log {S(0)}$ which increases
linearly with $T$ upto about $0.5J$. This is consistent with our crossover
expression, but inconsistent with the renormalized-classical formula,
(\ref{S5'})
which predicts that  $k_B  T \log {S(0)}$ decreases
with temperature.
 We therefore do not believe that the numerical
data correspond to the classical regime, as was suggested in Ref~\cite{Young}.

Finally, the correlation length.
 Series expansions reported that $\xi$
is approximately one lattice spacing at $k_B T =0.4 J$. This is
substantially lower than our renormalized-classical result
$\xi_{cl} \approx 5 a$ at the same temperature, but is consistent
with the value of $\xi$ deep
in the quantum-critical regime $\xi_{quant} \approx 1.25 a$. We
emphasize
however  that Ref.~\cite{Young} defined $\xi$ as $\xi^2 = - (1/S(k))
(\partial S/\partial k^2 )|_{k=0}$ - this definition yields a
nonuniversal value of $\xi$ for quantum-critical frustrated antiferromagnets.
On the
contrary,  our definition of $\xi$, from the long-distance decay of
spin-spin correlator, always yields a universal result.
Besides, even if the universal piece in $S(k)$ is dominant, as in the
renormalized classical regime,
the two definitions are still nonequivalent even at $N = \infty$ simply because
spin structure factor is related to the polarization operator of $z-$fields,
which unlike $z-$field propagator, does not have a Lorentzian form. In the
classical regime, the rescaling factor between the two definitions of $\xi$
 is  $\xi^{2}_{\rm series} = (2/3) \xi^{2}_{\rm ours}$ at $N = \infty$. The
value of
the rescaling factor in the intermediate and quantum-critical
 regimes is difficult to estimate, but on general grounds it should be
smaller than $2/3$ because  the nonuniversal piece in $S(0)$ becomes
 dominant at $g = g_c$. We therefore expect that the actual correlation length
is in fact larger than reported in~\cite{Young}. This again is consistent with
our observation that at $k_B T$ around $0.4J$, the system is in the crossover
region between renormalized-classical and quantum-critical regimes, and is
probably
closer at $k_B T = 0.4J$ to the quantum-critical regime.

\section{conclusions}
\label{conl}
In conclusion, we summarize  our main results.
 We have presented a general scaling framework to describe frustrated
antiferromagnetic systems near the quantum phase transition between classically
ordered and quantum-disordered ground states.

We considered various scaling functions for experimentally measurable
quantities both on the ordered and disordered sides of the quantum transition
and have shown that the observables which probe the behavior of
antiferromagnets at low energies are completely universal functions of
just a few measurable parameters at $T=0$.
On the ordered side, these parameters
are sublattice magnetization and
transverse and longitudinal spin-stiffness and spin susceptibility.

We then specialized to particular field-theoretic model of the transition
(results
for a different model are briefly noted in Appendix~\ref{confined})
Our approach began with the
fundamental assumption
 that the disordering transition at $T=0$ is continuous and that
 vortex-like excitations with a nonzero local $Z_2$ flux are
irrelevant at low energies. We showed that, in this situation,
 the proper low-energy theory near the transition is given by the
$SU(2) \times U(1)$ sigma-model for  spinon fields. All physically observable
excitations are collective modes of two
spinons. The global $SU(2)$ symmetry of the sigma-model action is related to
spin
rotations, while the global $U(1)$ symmetry is related to lattice
transformations~\cite{Aza}. For triangular and other commensurate noncollinear
antiferromagnets, this lattice symmetry in fact reduces to a discrete symmetry
($Z_3$
symmetry in case of triangular antiferromagnets). At the quantum transition
point,
the symmetry of the action enlarges to $O(4)$.

We then extended our action to a general $N$ by considering spinons as
$N$-component objects, and used the powerful technique of $1/N$ expansion.
 The extended action  has $SU(N) \times U(1)$ symmetry. The fixed point in
this approach has its
internal symmetry enlarged from $SU(N) \times U(1)$ to $O(2N)$
for {\it any} $N$.

We then
 used the $1/N$ expansion to explicitly compute the scaling properties of the
field-theory, always finding that they were
consistent with the more general scaling
ansatzes. We made definite predictions
for the dynamic structure
factor, static susceptibility, correlation length,
local and static structure factors, and the spin-lattice relaxation rate in the
renormalized-classical and  quantum-critical regions. We also briefly discussed
the low-$T$ behavior in the quantum-disordered region.

Finally, we compared our results to the properties $S=1/2$ triangular
antiferromagnets. We determined the input parameters in the scaling function
from
a $1/S$ expansion on the original lattice
Hamiltonian, and made quantitative predictions about the form of uniform
susceptibility, correlation length and
static structure factor. We compared the
results with the data of recent high-temperature series expansions.
All of the data were consistent with the interpretation that there is a narrow
window of quantum-critical behavior just below the temperature at which the
uniform
susceptibility passes through its maximum.
However, more detailed numerical and experimental results are needed before any
definitive conclusions can be reached. We hope
that it will be possible to perform measurements in a $T$ range
 between the $3d$ ordering temperature and the temperature
where uniform  susceptibility has a maximum.
Our prediction is that in between
the two temperatures, the uniform susceptibility should follow our
formula for the quantum-critical regime.

\section{acknowledgements}

The research was supported by NSF Grant No. DMR-92 24290.
We are pleased to thank P.Azaria, B. Delamott,
P. Lecheminant, D. Mouhanna, and N. Read for
useful discussions and  communications

\appendix
\section{Field Theory with confined spinons}
\label{confined}

In the event there is continuous transition from the magnetically
ordered state to a quantum-disordered state with {\em confined\/} spinons,
we expect that it can be described by a continuum field theory of the
${\bf n}_1$ and ${\bf n}_2$ fields themselves.
All fields are now singlets under the $Z_2$ gauge symmetry
and the $Z_2$ vortices are permitted.
Such large-$M$ field theories
have been considered earlier by Kawamura~\cite{Kawamura} and
Azaria {\em et. al.\/}~\cite{Aza}. A potential problem with this approach
is that the results of the $D=2+\epsilon$ analysis~\cite{Azaria} are not
obviously consistent with the $D=4-\epsilon$ and large $M$
theories.
The universal properties of such
nearly-critical antiferromagnets are rather similar to those of the
collinear antiferromagnets considered in Ref.~\cite{CSY}. Therefore
we will be rather brief, as the analog of all the results in the body of
this paper
can be obtained by minor modifications
of those of Ref~\cite{CSY}.

We will consider the action
\begin{equation}
{\cal Z} = \int {\cal D}{\bf n}_1 {\cal D}{\bf n}_2
\exp\left[-\frac{1}{2} \int d^D x \left(p_{1,\mu}
\left((\partial_{\mu} {\bf n}_1)^2 +
(\partial_{\mu} {\bf n}_2)^2 \right)  + p_{2,\mu} \left({\bf n}_1
\partial_{\mu} {\bf n}_2 -
{\bf n}_2 \partial_{\mu} {\bf n}_1 \right)^2 + V({\bf n}_1, {\bf n}_2)
\right)\right]
\label{N1}
\end{equation}
where $p_{1,\mu} = \rho^{0}_{\perp, \mu}, ~p_{2,\mu} =
(\rho^{0}_{\parallel, \mu} - 2 \rho^{0}_{\perp, \mu})/4$.
The potential $V({\bf n}_1 , {\bf n}_2 )$ can either impose the
hard-spin constraints (in a $D=2+ \epsilon$ expansion)
\begin{equation}
{\bf n}_1^2 = {\bf n}_2^2 = 1~~~;~~~{\bf n}_1 \cdot {\bf n}_2 = 0
\label{constr}
\end{equation}
or the soft-spin potential (in a $D=4-\epsilon$ expansion)
\begin{equation}
V ({\bf n}_1, {\bf n}_2) = \frac{1}{2} r_0 \left({\bf n}^{2}_1 +
 {\bf n}^{2}_2 \right) + u_1 \left({\bf n}^{2}_1 + {\bf n}^{2}_2 \right)^2 +
 u_2 \left({\bf n}_1 \times {\bf n}_2 \right)^2
\label{N3}
\end{equation}
Kawamura~\cite{Kawamura} introduced a large $M$ expansion of (\ref{N1})
in which
the vectors ${\bf n}_1$, ${\bf n}_2$ are generalized to $M$-components;
the action then has a $O(M) \times O(2)$ symmetry. The relationships
betwen the large $M$, $\epsilon=D-2$, and $\epsilon=4-D$ expansions
have been discussed by Azaria {\em et. al.\/}~\cite{Aza}.

Here we will discuss some simple properties of the large $M$ expansion.
The results have striking differences from the large $N$ expansion
of this paper, in particular, the phase transition at $M >3$ belongs to the
universality class different from $O(M+1)$ model.
 Which of these two expansions is more appropriate for the
physical case $M=3$, $N=2$ is not quite clear, and numerical studies of
frustrated antiferromagnets will be
quite useful in this regard. The most obvious difference
is of course in the absence of spinons in the large $M$ theory.
The staggered susceptibility $\chi_s (k , \omega )$ now has
delta-function quasiparticle peaks, in contrast to the branch cuts
of the large $N$ theory. Differences also appear in the
behavior of the correlators of the conserved charges and currents.
A key property of the $1/M$ expansion is that the $p_{2,\mu}$ couplings
are irrelevant. This immediately implies that the universal
ratios of the stiffnesses (Eqn (\ref{defups})) obey $\Upsilon_{\rho}
= \Upsilon_{\chi}  = 2$ at $M=\infty$. We computed $\Upsilon$ to first
order in $1/M$ and obtained
\begin{equation}
\Upsilon_{\rho} = \Upsilon_{\chi} = 2 - \frac{26}{3 \pi^2 M}
\end{equation}
In performing the $1/M$ calculations, we introduced three Lagrange multipliers
in
the functional integral to impose the constraints (\ref{constr}), and also
introduced condensates of the $n_1$ and $n_2$ fields. The computations are
a bit tricky: we found that the $1/M$ correction to $\Upsilon$ is
related to the difference in the Green functions of the transverse components
of $n_{1,2}$ and the fluctuating components along the directions of the
condensates. This difference clearly disappears at $g=g_{c}$; however the
correction to the ratio of the stiffnesses (i.e., to to $O(k^2)$ in the
full Green functions) remains finite at $g = g_c$
because it includes integrals which are divergent at $g \rightarrow
g_c$.

Also interesting is the behavior of the uniform susceptibility
$\chi_u (T)$ in the quantum-critical region. It is simple to show
that at $M=\infty$ this is given by precisely {\em twice} the
mean-field result of Ref~\cite{CSY}, with $c \rightarrow c_{\perp}$.
Contrast this with the result of the deconfined spinon model of the
body of the paper: there we found that the $N=\infty$ result
was {\em one-half\/} the result of Ref~\cite{CSY} !

\section{Spin-wave calculations at $T=0$}
\label{largeS}
For experimental comparisons of the results obtained within $1/N$ expansion,
we need the $T=0$ expressions for  sublattice magnetization, spin-wave
velocities and uniform spin susceptibilities. Below we will calculate
these quantities for the Heisenberg antiferromagnet on a triangular lattice
in an expansion in $1/2S$,
where $S$ is the value of the spin. Though we will use large $S$ approach,
 our chief interest is  in the case of $S=1/2$ when
quantum fluctuations
are the strongest.
As we will see below,
 the convergence of the perturbative
series in $1/2S$ in triangular
antiferromagnets is very good (as it is on the square
lattice~\cite{Girvin_Canali,Igarashi}), and the $1/S$ expansion is likely
 to give quite accurate values of observables, even for
$S=1/2$.

We now turn to a description of the calculations. We consider here the model
 with interactions between nearest neighbors:
\begin{equation}
{\cal H} = J \sum_{\bf{l},\bf{\Delta}} \bf{S}_{\bf{l}}
\bf{S}_{\bf{l}+\bf{\Delta}}.
\label{A1}
\end{equation}

The procedure  of doing the $1/S$ expansion is rather standard and involves
several steps which include (i) the transformation
from spin operators to bosons via
Holstein-Primakoff, Dyson-Maleev, or some other transformation, (ii)
the diagonalization of the quadratic form in bosons,
and  (iii) the use of a standard perturbative technique for
Bose-liquids to
treat the interaction between spin waves.
Noninteracting spin waves have energy which scales as
 S, while the
interaction vertex involving  $m$ bosons scales as $ S^{2-m/2}$;
this gives rise to an expansion in powers of $1/S$ for anharmonic
contributions, similar to that in a weakly interacting Bose gas.

Another important issue related to the $1/S$ expansion,
 is the number of bose fields which one has to introduce in order to
keep track of  the whole spin-wave
spectrum, not just the low-energy modes. This is important because quantum
fluctuations are not divergent in $2d$, and the $1/S$
expansion  involves sums over the whole Brillouin zone.
In the general case, the number of
different bose fields is equivalent to
the number of magnetic sublattices. However, in several special cases,
a multisublattice  magnetic configuration can be transformed into a
one-sublattice  ferromagnetic one by applying
a uniform twist on the  coordinate frame.
In this situation,  the spin-wave spectrum has no gaps at the
boundaries of the reduced Brillouin zone and
one can describe all excitations by a single bosonic field,
 as in the case of a  ferromagnet.
 Obviously, the triangular antiferromagnet in a zero magnetic field
is an example of such special behavior: the $120^o$ ordering becomes a
ferromagnetic one in the twisted coordinate frame with a pitch $Q=(4\pi/3, 4\pi
/\sqrt{3})$. We therefore will use a one-sublattice description of triangular
antiferromagnet whenever possible.  This indeed substantially simplifies the
calculations.

We start with the transformation from spin operators to
bosons. The  choice of the transformation
is  indeed only a matter of convenience,
and the final results are  independent of the way how bosons are
introduced. Nevertheless, there  are several possibilities extensively
discussed in the literature~\cite{Kag_Chub}. We found it most
convenient to use here the
conventional Holstein-Primakoff transformation because it preserves
the Hermitian properties of the Hamiltonian.
We therefore use
\begin{equation}
S_z = S - a^{\dagger}a; ~~
S^{+} = \sqrt{2 S - a^{\dagger}a}~~ a; ~~
S^{-} =  a^{\dagger}  \sqrt{ 2 S - a^{\dagger}a}
\label{HP}
\end{equation}
Substituting this transformation into (\ref{A1}), expanding the radical,
and restricting to only cubic and quartic anharmonic terms, we obtain
after some algebra
\begin{equation}
{\cal H} = {\cal H}_0 + 3 J S ({\cal H}_2 + {\cal H}_3 + {\cal H}_4)
\label{A2}
\end{equation}
where ${\cal H}_0 = -\frac{3}{2} J S^2 N$ is the classical ground state
energy, and other terms are
\begin{eqnarray*}
{\cal H}_2 &=& \sum_k A_k a^{\dagger}_k a_k + \frac{B_k}{2} (a^{\dagger}_k
a^{\dagger}_{-k} + a_{k} a_{-k}) \\
{\cal H}_4 &=& \frac{1}{16 S} \sum a^{\dagger}_1
a^{\dagger}_2 a_3 a_4 \left[4 (\nu_{1-3} +\nu_{2-3}) + \nu_1 +\nu_2 +\nu_3
+\nu_4 \right]\\
&& - 2 \left(a^{\dagger}_1 a^{\dagger}_2 a^{\dagger}_3 a_4 +
a^{\dagger}_4 a_3 a_2 a_1 \right) (\nu_1 + \nu_2 +\nu_3)
\end{eqnarray*}
\begin{equation}
{\cal H}_3 =  i \sqrt{\frac{3}{8 S}} \sum (a^{\dagger}_1 a^{\dagger}_2 a_3 -
a^{\dagger}_3 a_2 a_1) (\bar{\nu}_1 + \bar{\nu}_2).
\label{A3}
\end{equation}
Here $i \equiv k_i$, and
\begin{equation}
\nu_k = \frac{1}{3} \left(\cos{k_x} + 2 \cos{\frac{k_x}{2}} \cos{\frac{k_y
\sqrt{3}}{2}} \right); ~~
\bar{\nu}_k = \frac{2}{3} \sin{\frac{k_x}{2}} \left(\cos{\frac{k_x}{2}} -
\cos{\frac{k_y \sqrt{3}}{2}} \right).
\label{A4}
\end{equation}
Finally, $A_k$ and $B_k$ are given by
\begin{equation}
A_k = 1 + \frac{\nu_k}{2};~~~~~~B_k = -\frac{3}{2} \nu_k
\label{A6}
\end{equation}
At $S=\infty$, anharmonic terms are absent and ${\cal H}_1$ can be
diagonalized by a standard Bogolubov transformation
\begin{equation}
a_k = l_k(c_k + x_k c^{\dagger}_{-k})
\label{A7}
\end{equation}
with
\begin{equation}
l_k = \left(\frac{A_k + E_k}{2 E_k}\right)^{1/2}; ~~ x_k = -\frac{B_k}{|B_k|}
\left(\frac{A_k - E_k}{A_k + E_k}\right)^{1/2}.
\label{A8}
\end{equation}
and
\begin{equation}
E_k = (A^2 _k - B^2 _k)^{1/2} = \left((1-\nu_k)(1 +2 \nu_k)\right)^{1/2}
\label{A9}
\end{equation}
The diagonalization yields
\begin{equation}
{\cal H}_1 = \sum_k E_k c^{\dagger}_k c_k
\label{A10}
\end{equation}
It follows from Eqn. (\ref{A9}) that the excitation spectrum
of the ideal gas of magnons has three zero modes, as it indeed should.
 Two of these modes are at $k = \pm Q$
where $Q = (4\pi /3, 4\pi /\sqrt{3})$ is the ordering momentum in triangular
antiferromagnet, and the third is at $k=0$ and describes soft fluctuations of
 total magnetization.
 The expansion near zero modes gives two  spin-wave
velocities
\begin{equation}
c_{\perp} = c_{\pm Q} = \frac{3 \sqrt{3}}{2 \sqrt{2}} JS a
\end{equation}
\begin{equation}
c_{\parallel} = c_{k=0} = \frac{3 \sqrt{3}}{2} JS a
\label{A11}
\end{equation}
The ratio of the two at $S = \infty$ is
$c_{\parallel}/ c{\perp} = \sqrt{2}$. This was also obtained in
other approaches~\cite{Dombre_Read}.

The infinite $S$ spin-wave results can be also used to get the first
quantum correction to on-site magnetization~\cite{Jolicoeur}.
Indeed, $\langle a^{\dagger} a \rangle$ in
(\ref{HP}) is nothing but the  density of particles which is finite due to
the anomalous term in the quadratic form.
{}From (\ref{A7}, \ref{A8}), we have
$\langle a^{\dagger}_k a_k \rangle = (A_k - E_k)/ 2 E_k$, and
therefore noninteracting
spin waves reduce the sublattice magnetization to
\begin{equation}
<S> = S \left(1 - \frac{1}{2S} ~\sum _k \frac{A_k - E_k}{E_k} \right) =
 S \left(1 - \frac{0.522}{2S} \right)
\label{A12}
\end{equation}

We next consider corrections to Eqns (\ref{A10})
and (\ref{A11}) due to the interactions between spin-waves.
We will follow the same line of reasoning as
for square-lattice antiferromagnets. However, the presence of cubic
terms makes the analysis considerably more involved.

We start with the spin-wave velocity renormalization.

\subsection{Spin-wave velocity}

Our goal is to obtain the leading $1/S$ renormalization of
 spin-wave excitations. For this we consider  first-order self-energy
corrections due to quartic anharmonicities
and second-order corrections due to cubic anharmonicities (recall that cubic
terms have the overall factor $S^{1/2}$). The corrections due to quartic
terms are easy to compute, because to leading order in $1/S$, one can get
away
with simple one-loop diagrams. Equivalently, one can
simply decouple the four-fold term in eq. (\ref{A3}) by making all possible
pair averaging. The
quadratic form  allows for nonzero normal $\langle a^{\dagger}_k a_k \rangle$
and
anomalous $\langle a_k a_{-k} \rangle $ pair products of Bose particles, and
the
decoupling
 changes $A_k$ and $B_k$ to
\begin{equation}
\bar{A}_k = \left(1 + \frac{\nu_k}{2} \right) \left( 1 + \frac{1}{2S}  -
\frac{1}{2S}~ \sum_p \frac{1}{E_p} \left(1 + \frac{\nu_p}{4} + \nu^{2}_p
\right) \right) - \frac{3}{8S} \sum_p \frac{\nu_p}{E_p} (1 - 4 \nu_p)
\label{A13}
\end{equation}
\begin{equation}
\bar{B}_k = - \frac{3}{2} \nu_k \left(1 + \frac{1}{2S} - \frac{1}{2S} \sum_p
\frac{1}{E_p}~\left(1 + \frac{\nu_p}{4} - \nu^{2}_p \right)\right) +
\frac{3}{8S} \sum_p \frac{\nu_p}{E_p}
\label{A14}
\end{equation}
A simple inspection then shows that the renormalized spectrum ($\bar{E}_k =
(\bar{A}^{2}_k - \bar{B}^{2}_k)^{1/2}$) still keeps a zero mode at $k=0$,
but acquires a finite gap  at $k = \pm Q$:
\begin{equation}
E^{2}_Q = - \frac{9}{8S}~ \sum_p \frac{\nu_p (1- \nu_p)}{E_p}
\label{AA1}
\end{equation}
 This finite gap is indeed an
artifact of using only quartic terms, and cubic anharmonicities
should restore the correct structure of the spectrum,
as we demonstrate below.

There are several ways to deal with the cubic terms: one can either
calculate the effective four-fold vertex produced by two triple
vertices~\cite{Rast_Tassi,Chub},
and then use the decoupling procedure, or one can transform to quasiparticles
(i.e., diagonalize the quadratic form)
considering first only quartic corrections, and then
calculate the renormalization of the excitation spectrum due to cubic terms
in the second-order perturbation theory.
Below we use the second approach which is technically
advantageous. We therefore first transform from particle operators ($a_k$)
to quasiparticles ($c_k$) using eq. (\ref{A7}), but with $\bar{A}_k$ and
$\bar{B}_k$ instead of $A_k$ and $B_k$. The bare Hamiltonian then keeps
 the form of eq.(\ref{A10})  with $\bar{E}_k$ instead of $E_k$.
On the other hand, the structure of cubic vertices becomes more involved
 after the transformation to
quasiparticles, and instead of  Eqn. (\ref{A3}) we obtain
\begin{equation}
{\cal H}_3 = i \sqrt{\frac{3}{32S}}~ \sum~
 c^{\dagger}_1 c^{\dagger}_2 c_3 \Phi_1 (1,2;3) ~+~ \frac{1}{3}
c^{\dagger}_1 c^{\dagger}_2
c^{\dagger}_3 \Phi_2 (1,2,3)  + H.c
\label{A15}
\end{equation}
The vertex functions $\Phi_1$ and $\Phi_2$ are given by
\begin{equation}
\Phi_1 (1,2;3) =
\frac{\tilde{\Phi}_1 (1,2;3)}{\sqrt{\bar{E}_1 \bar{E}_2 \bar{E}_3}}; ~~~
\Phi_2 (1,2,3) =
\frac{\tilde{\Phi}_2 (1,2,3)}{\sqrt{\bar{E}_1 \bar{E}_2 \bar{E}_3}}
\label{A16}
\end{equation}
where
\begin{eqnarray*}
\tilde{\Phi}_1 (1,2;3) &=&
 \bar{\nu}_1 f^{(1)}_{-} (f^{(2)}_{+} f^{(3)}_{+} +
f^{(2)}_{-} f^{(3)}_{-}) + \bar{\nu}_2 f^{(2)}_{-} (f^{(1)}_{+} f^{(3)}_{+} +
f^{(1)}_{-} f^{(3)}_{-}) + \bar{\nu}_3 f^{(3)}_{-} (f^{(1)}_{+} f^{(2)}_{+} -
f^{(1)}_{-} f^{(2)}_{-})\\
\tilde{\Phi}_2 (1,2,3) &=&
\bar{\nu}_1 f^{(1)}_{-} (f^{(2)}_{+} f^{(3)}_{+} -
f^{(2)}_{-} f^{(3)}_{-}) + \bar{\nu}_2 f^{(2)}_{-} (f^{(1)}_{+} f^{(3)}_{+} -
f^{(1)}_{-} f^{(3)}_{-}) + \bar{\nu}_3 f^{(3)}_{-} (f^{(1)}_{+} f^{(2)}_{+} -
f^{(1)}_{-} f^{(2)}_{-})
\end{eqnarray*}
and
\begin{equation}
f^{(i)}_{\pm} = (\bar{A}_i \pm \bar{B}_i)^{1/2}  .
\label{A18}
\end{equation}
The self-energy diagrams to order $1/S$ are shown on
Fig.~\ref{sec_or_dia}. We see that cubic terms give rise to
 both normal and anomalous
self-energy parts  so that  the dispersion relation
again has the form typical for a $2\times2$ problem:
\begin{equation}
(\omega +  \Sigma_a (k, \omega) )^2 = (\bar{E}_k + \Sigma_s (k, \omega))^2
- (\Sigma_{+,+} (k, \omega))^2
\label{A19}
\end{equation}
where $\Sigma_{s,a} (k, \omega) = \frac{1}{2}~(\Sigma_{+,-} (k, \omega)
\pm \Sigma_{+,-} (-k, -\omega)$. However, it is not difficult to check that
 $\Sigma_{-,-} \sim \Sigma_{+,+} \sim 1/S$ and therefore
anomalous self-energy terms  contribute to the excitation energy only to
order $1/S^2$, while to order $1/S$ a solution of Eqn (\ref{A19}) is
simply $\omega = \tilde{E}_k$ where
\begin{equation}
\tilde{E}^{2}_k = \bar{E}^{2}_k + 2 \bar{E}_k \Sigma_{+,-} (k, \bar{E}_k)
\label{A20}
\end{equation}
We therefore need to evaluate here only the normal component of
the self-energy. The analytical expression for $\Sigma_{+,-}$ is
\begin{equation}
 \Sigma_{+,-} (k, \bar{E}_k) = - \frac{3}{16S}~ \sum
\left (\frac{|\Phi_1 (1,2;k)|^2}{\bar{E}_1 + \bar{E}_2 - \bar{E}_k} +
\frac{|\Phi_2 (1,2,k)|^2}{\bar{E}_1 + \bar{E}_2 + \bar{E}_k} \right)
\label{A21}
\end{equation}
 To leading order in $1/S$ we can indeed use nonrenormalized values for
$A_k, B_k, E_k$  in the r.h.s. of (\ref{A21}).

We first demonstrate that  $\tilde{E}_k$ has a true zero mode at $k=Q$.
For this we need to evaluate $\Sigma_{+,-} (Q, \bar{E}_Q)$. We found
the following equality to be quite useful in the calculation
\begin{equation}
\sqrt{3}~\bar{\nu}_{q \pm Q/2} = (A_{q \pm Q/2} + B_{q \pm Q/2}) -
(A_{q \mp Q/2} - B_{q \mp Q/2})
\label{A22}
\end{equation}
Substituting (\ref{A22}) into the expressions for the vertex
functions and using  $A_Q = B_Q = 3/4$, we obtain after
simple algebra
\begin{equation}
\tilde{\Phi}_1 (1,2; Q) =  \tilde{\Phi}_2 (1,2,Q) =
 \frac{(E_1 + E_2)}{\sqrt{2}} ~~
 (f^{(1)}_{+} f^{(2)}_{+} - f^{(1)}_{-} f^{(2)}_{-})
\label{A23}
\end{equation}
Substituting, then, the vertex functions into the formula for the
self-energy we obtain using (\ref{A18})
\begin{equation}
\Sigma_{+,-} (Q, E_Q) =  \frac{1}{2E_p} ~\frac{9}{8S}
 \sum_p \frac{\nu_p (1 - \nu_p)}{E_p}
\label{A24}
\end{equation}
Finally, upon substituting  this result into Eqn (\ref{A20}) and
using (\ref{AA1}) for $\bar{E}_Q$, we find that the gap in the excitation
spectrum disappears as it should~\cite{comment}.

Our next step is to expand $\bar{E}_k$ and $\Sigma_k$ near the zero modes,
and obtain the corrections to the spin-wave velocities to order $1/S$.
The expansion near $k =0$ is quite straightforward because
$\tilde{\Phi}_1 (1,2;k)$ and
$\tilde{\Phi}_2 (1,2,k)$ both scale as $k$ at small $k$, and one can
therefore safely neglect $E_k$ in the denominators in (\ref{A21}).
 Doing the algebra, we obtain
the renormalized spin-wave velocity at $k\approx 0$ in the form
\begin{equation}
\tilde{c}_{\parallel} = c_{\parallel} \left(1 + \frac{1}{2S} - \frac{1}{3S}~
\sum \frac{Q^{2}_k}{E_k} ~\left(\frac{5}{2} - Q^{2}_k + \frac{3}{8}~(9 -
4Q^{2}_k)~\Lambda_k \right)\right)
\label{A25}
\end{equation}
where
\begin{eqnarray*}
 Q^{2}_k &=& \sin^{2}\frac{k_x}{2} +
 \sin^{2} \frac{k_x + \sqrt{3} k_y}{4} +
\sin^{2} \frac{k_x - \sqrt{3} k_y}{4}\\
\Lambda_k &=& -\frac{1}{3} +
{}~\left(\sin^{4}\frac{k_x}{2} ~+~
\sin^{4} \frac{k_x + \sqrt{3} k_y}{4} ~+~
\sin^{4} \frac{k_x - \sqrt{3} k_y}{4}\right)/ Q^{4}_k
\end{eqnarray*}
Numerical intergation then gives
\begin{equation}
\tilde{c}_{\parallel} = c_{\parallel} \left(1 - \frac{0.115}{2S} \right)
\label{A26}
\end{equation}
The structure of the expansion near $k = \pm Q$ is more involved and
we refrain from presenting the analytical  expression for the spin-wave
velocity. Numerically, we obtained
\begin{equation}
\tilde{c}_{\perp} = c_{\perp} \left(1 + \frac{0.083}{2S} \right)
\label{A27}
\end{equation}

Comparing (\ref{A26}) and (\ref{A27}), we observe
 that  quantum fluctuations tend to
diminish the difference between the two spin-wave velocities. This is
 consistent with our  result (\ref{C14'}) that the relative difference
between $\tilde{c}_{\perp}$ and $\tilde{c}_{\parallel}$ should disappear
 at the quantum-critical point.
We will use (\ref{A26}) and (\ref{A27})
 below and now proceed with the calculations of sublattice magnetization.

\subsection{Sublattice magnetization}

We have shown above that to leading order in $1/S$,
the correction to sublattice magnetization
comes already from noninteracting magnons (Eqn (\ref{A12})). Here we
obtain the next term in the expansion in $1/S$,  which is also the leading
$1/S$ correction to the density of particles.  We again have to consider
both quartic and cubic terms, since they contribute at the same order to
$\sum_k  \langle a^{\dagger}_k a_k\rangle$.
As before, quartic terms only renormalize the coefficients in
the quadratic form,
and hence change the expression for the density of particles to
\begin{equation}
\sum_k  \langle a^{\dagger}_k a_k \rangle ~=~ -\frac{1}{2} + \frac{1}{2} \sum_k
{}~\frac{\bar{A}_k}{\bar{E}_k}
\label{A28}
\end{equation}
where $\bar{A}_k$ and $\bar{E}_k$  are given by ({\ref{A13}) and ({\ref{A14}).
In explicit form
\begin{equation}
\frac{1}{2} \sum_k \frac{\bar{A}_k}{\bar{E}_k} = \frac{1}{2} \sum_k \frac{1 +
{}~\nu_k /2}{E_k} ~-~ \frac{9}{32 S}~ \sum_p \frac{\nu^{2}_p}{E_p}~\sum_q
\frac{\nu_q}{E_q}  ~-~ \frac{9}{32 S}~ \sum_p
\frac{\nu_p (1- \nu_p)}{E_p}~\sum_q
\frac{\nu_q (1 - \nu_q)}{E^{3}_q}
\label{A29}
\end{equation}
We see that the very last term behaves near $q = Q$ as $|q -Q|^{-3}$ which
makes the integral over $q$ divergent.
The divergence is indeed an artificial one and should disappear
when we add the contributions
of the cubic terms.

To see how cubic terms modify (\ref{A28}), we express the density
of particles in terms of the quasiparticles operators using (\ref{A7}) and
(\ref{A8}):
\begin{equation}
\sum_k  \langle a^{\dagger}_k a_k \rangle ~=~ -\frac{1}{2} + \frac{1}{2}
\sum_k            \frac{\bar{A}_k}{\bar{E}_k} ~-~ \sum_k
\frac{\bar{B}_k}{\bar{E}_k}~<c_k c_{-k}> ~+~
\sum_k \frac{\bar{A}_k}{\bar{E}_k}~<c^{\dagger}_k c_{k}>
\label{A30}
\end{equation}
The first two terms are just the renormalized spin-wave terms. The third
correction is related to  the anomalous
self-energy term in Fig.\ref{sec_or_dia}. Performing
the frequency summation in this term, we obtain
\begin{equation}
{}~-~ \sum_k \frac{\bar{B}_k}{\bar{E}_k}~<c_k c_{-k}> = -\frac{9}{32S}~
\sum_k \frac{\nu_k}{E^{3}_k}~\Psi_k
\label{A31}
\end{equation}
where
\begin{equation}
\Psi_k =  \sum ~\frac{1}{E_1 E_2}~
\frac{\tilde{\Phi}_1 (1,2;k) \tilde{\Phi}_2 (1,2, -k)}{E_1 + E_2 +
 E_k}
\label{A32}
\end{equation}
Finally, the last term in (\ref{A30}) contains the density
of quasiparticles. This density is finite to order $1/S$ because
among cubic non-linearities, there is the term which describes
 simultaneous emission  of three spin-waves. Evaluating the expectation
value of $\langle c^{\dagger}_k c_{k} \rangle$ by the usual means, we obtain
\begin{equation}
 \sum_k \frac{\bar{A}_k}{\bar{E}_k}~<c^{\dagger}_k c_{k}> = \frac{3}{16 S}~
\sum_k \frac{1 +
{}~\nu_k /2}{E^{2}_k}~\Upsilon_k
\label{A33}
\end{equation}
where
\begin{equation}
\Upsilon_k =  \sum ~\frac{1}{E_1 E_2}~
\frac{|\tilde{\Phi}_2 (1,2, k)|^2}{(E_1 + E_2 +
 E_k)^2}
\label{A34}
\end{equation}

We first show  that the total expression for the density of particles
is free from divergencies. Simple inspection of Eqns (\ref{A31})
 - (\ref{A34})
shows that the divergent contributions from the cubic terms (namely, $1/E^3$
and $1/E^2$ terms in (\ref{A31}) and $1/E^2$ terms in (\ref{A33}) )
come from the region $k \approx Q$, where $\Psi$ and $\Upsilon$ tend to
constant values. For these  $k$, we again use  (\ref{A22}),
substitute it into the vertex functions, and after a
simple algebra obtain
\begin{equation}
\Psi_k = - (1 - \nu_Q)~ \sum_p \frac{\nu_p (1 - \nu_p)}{E_p} - E_k \Upsilon_Q +
O(E^{2}_k)
\label{A35}
\end{equation}
Substituting further this expression into (\ref{A31}) and
comparing the result with the divergent piece in (\ref{A29}),
we find that the $1/E^3$ contributions from
cubic and quartic terms, and the $1/E^2$ contributions from the two cubic terms
cancel each other, so that the
  $1/S$ correction to
the density of particles is finite, as it of course should be.
We then performed numerical  computation of the $1/S$ terms in (\ref{A30})
and obtained
\begin{equation}
\langle S \rangle =  \left(S - 0.261 + \frac{0.027}{(2S)} \right)
\label{A36}
\end{equation}
For $S=1/2$, Eqn (\ref{A36}) yields $\langle S \rangle \approx 0.266$, which is
close
to half of the classical value. A very
similar result was obtained earlier by Miyake~\cite{Miyake}, who calculated
the on-site magnetization to order $1/S^2$ by
evaluating numerically the response to a staggered magnetic field. His
estimate for the $1/S^2$ correction
is  however somewhat smaller than ours ($0.01$ instead of
$0.027$). In any event, $1/S^2$ terms are rather small and can hardly change
substantially the lowest-order spin-wave result for the
magnetization~\cite{Jolicoeur}.
We therefore found no support
for the recent claim~\cite{Singh-Huse} that the value of magnetization is
substantially lower than the spin-wave prediction.
 Note, in passing, that for square lattice antiferromagnet,
 the first anharmonic correction to $\langle S \rangle$ is
identically zero~\cite{Chak_Cast}. Indeed, cubic terms are absent in the
square-lattice antiferromagnet, and
$1/S$ corrections  due to quartic terms
do not change the shape of the quasiparticle spectrum (that is, $\bar{A}_k
/\bar{E}_k ~=~A_k /E_k$ ).  The next to leading order correction
in the square-lattice case has been calculated and found to be very
small~\cite{Chak_Cast}.

\subsection{Uniform susceptibility}

Now we calculate, to order $1/S$, the response of a triangular antiferromagnet
 to an external magnetic field. We have already discussed in
Sec.\ref{cons_curr}
 that the  magnetic susceptibility  tensor in a
triangular antiferromagnet has the form~\cite{AM}
\begin{equation}
\chi_{\alpha ~\beta} = \chi_{\perp} \delta_{\alpha\beta} +
(\chi_{\parallel} - \chi_{\perp})
 m_{\alpha} m_{\beta}
\label{A37}
\end{equation}
where $\bf{m}$ is a unit vector which specifies the plane of spin ordering.
 This form of $\chi_{\alpha\beta}$ implies
 that the ordered
state should  have two different spin susceptibilities.
 They can be viewed
as the response to the field applied perpendicular to the plane of spin
ordering,
i.e., along ${\vec  m}$
($\chi_{\parallel}$), and as the response  to a field directed in the plane
($\chi_{\perp}$). In the latter case, we need to introduce an
infinitesimally small anisotropy which keeps the spins in the basal plane.

For classical spins,
the transverse and longitudinal susceptibilities can easily be obtained
 by minimizing the ground state energy.
This yields $\chi_{\perp} = \chi_{\parallel} = 2 /9 \sqrt{3} J a^2$
where $a$ is the
interatomic spacing ($a^2 \sqrt{3}/2$ is the unit cell volume). As in
the bulk of the paper, we define $\chi_{\perp}$ and $\chi_{\parallel}$ without
the gyromagnetic ratio  $g \mu_B /\hbar$. We see that
the two susceptibilities are equal in the
classical limit~\cite{Kaw_Miyash}. This
 degeneracy in the response to a magnetic field in a 2D triangular
antiferromagnet has attracted some attention in the past as an example
of the "order from disorder" phenomenon~
\cite{Kaw_Miyash,Korsh,Muller_Hartm,Chub_Gol,Henley}. For our present purposes,
it is
sufficient to observe that the degeneracy is a purely classical effect. It
is not related to the symmetry properties of a quantum system and
 therefore should be broken by quantum fluctuations.

Technically, the computations in a finite field are
more involved because the transverse field
 breaks the $120^o$ ordering in the basal plane. In this case,
 a transformation to a twisted coordinate frame is no longer  advantageous
 because umklapp processes also contribute to order $1/S$.
 It is then more convenient to introduce a
separate bose field for each of three sublattices. For the longitudinal
 response, the $120^o$ ordering in the basal plane is preserved and
 a one-sublattice description with no umklapp terms is still valid. However,
one has to be careful  in this case as well, because
in the presence of a field, the excitation spectrum
 is no longer an even function of $k$. This is consistent with the fact that
 time reversal symmetry in a magnetic field requires that
in changing $k \rightarrow -k$ in the spectrum,
 one  has to change simultaneously the sign of $H$.

The corrections to the susceptibility tensor to order $1/S$ were computed
by Golosov and one of us~\cite{Chub_Gol}. We refrain from presenting the
details of the calculations and list here only the
 results. To order $1/S$, they are (notice that the definitions of
$\chi_{\perp}$ and $\chi_{\parallel}$ in~\cite{Chub_Gol} are interchanged
compared to ours):
\begin{equation}
\chi_{\perp} = \frac{2}{9 \sqrt{3} J a^2}~ Z^{\chi}_{\perp}; ~~
\chi_{\parallel} = \frac{2}{9 \sqrt{3} J a^2}~ Z^{\chi}_{\parallel}
\label{A38}
\end{equation}
where
\begin{equation}
 Z^{\chi}_{\parallel} = \left(1 - \frac{1}{2S}~\sum_k \frac{\nu_k
(1 - \nu_k)}{E_k} \right) = 1 - \frac{0.448}{2S}.
\label{A39}
\end{equation}
and
\begin{equation}
 Z^{\chi}_{\perp} = \left(1 - \frac{1}{2S}~\sum_k \frac{\nu_k
(1 + 2\nu_k)}{E_k} +  \frac{3}{2S}~\sum_k ~\frac{\nu^{2}_k}{E_{k_1} + E_{k_2}}
{}~\frac{f^{(1)}_{-}~f^{(2)}_{-}}{E_{k_1} E_{k_2}} \right) =
1 - \frac{0.291}{2S}.
\label{A40}
\end{equation}
where $f^{(i)} \equiv f^{(k_i)}$ were defined in (\ref{A18}).
Note that contrary to the situation in a stacked
$3d$ triangular antiferromagnet
 where $\chi_{\parallel} > \chi_{\perp}$,
the transverse (in-plane) susceptibility in the $2d$
case turns out to be larger than the longitudinal one;  this gives rise to
an unconventional phase diagram in a magnetic field which
has been discussed several times in the
literature~\cite{Kaw_Miyash,Korsh,Muller_Hartm,Chub_Gol}.

\subsection {Spin stiffness}
With the values of the two spin-wave velocities and
spin susceptibilities at hand, we are now in a position to calculate the spin
stiffnesses. To order $1/S$ they are
\begin{equation}
\rho_{\perp} = \chi_{\perp} c^{2}_{\perp} =
\frac{\sqrt{3}}{4} J S^2 Z^{\rho}_{\perp} ~~~
\rho_{\parallel} = \chi_{\parallel} c^{2}_{\parallel} =
\frac{\sqrt{3}}{2} J S^2 Z^{\rho}_{\parallel}
\label{A41}
\end{equation}
where
\begin{equation}
Z^{\rho}_{\perp} = 1 - 0.125/2S, ~~  Z^{\rho}_{\parallel} = 1 - 0.678/2S
\end{equation}

Finally, substituting the results (\ref{A38})- (\ref{A41}) into (\ref{C14}),
we ontain for $N=2$
\begin{equation}
\rho_s = \frac{1 - 0.402/2S}{\sqrt{3}}~JS^2, ~~~~\chi =
\frac{2}{9 \sqrt{3} J a^2}~ \left(1 - 0.343/2S \right)
\label{A42}
\end{equation}

\section{Computations in the N\'{e}el state at $T=0$}
\label{appendt=0}
In this Appendix, we derive to order $1/N$, the expression for $T=0$ sublattice
magnetization  near the quantum phase transition.
This result will be used in the derivation
of the universal scaling forms for uniform and staggered susceptibilities
in both renormalized-classical and quantum-critical regions. We also
reproduce the $1/N$ expressions for the relative difference between
longitudinal
and transverse spin-stiffness and spin susceptibility which were obtained by
other means in Sec~\ref{cons_curr}.

 Our  point of departure is the functional integral for
the $SU(N) \times U(1)$ $\sigma-$model, Eqn (\ref{C1}).
At $T=0$, the spin-rotation
symmetry is broken, and we represent the $N$ component complex
 vector $z$ of length $N$ as
\begin{equation}
z = (\sigma_0 + i \sigma_1, \pi_1 + i \pi_2, \pi_3 + i \pi_4 ...),
\label{B0}
\end{equation}
 where $ \langle \sigma_0\rangle$ is finite. This
parametrization  slightly differs from (\ref{Zord}), in that $\sigma_1$
and $\pi_i$ are defined without a factor $1/2$; notice also that
 we do not neglect
fluctuations in the direction of the condensate.
 Upon substituting (\ref{B0}) into (\ref{C1}),
 the functional integral becomes
\widetext
\begin{equation}
{\cal Z}   =  \int {\cal D} \sigma_{0}~ {\cal D} \sigma_{1}~
 {\cal D} \pi_{l} \,\,
\delta({\sigma}^{2}_{0} + {\sigma}^{2}_{1}  +  \pi^{2}_{l}
-N)~e^{- {\cal S}}
\end{equation}
where
\begin{eqnarray}
{\cal S} &=& \int d^2 r \int^{\infty}_{0}
d \tau ~ \sum_{\mu = {\vec x}, t} \sum_{l,m}
\frac{1}{g_{\mu}}~\left((\partial_{\mu}\sigma_{0})^{2}
+ (\partial_{\mu}\sigma_{0})^{2} + (\partial_{\mu} \pi_{l})^2 \right)
+\nonumber \\
&& \frac{\gamma_{\mu}}{N g_{\mu}} \left(\sigma_0 \partial_{\mu}
\sigma_1 - \sigma_1
\partial_{\mu} \sigma_0
\pi_{2m} \partial_{\mu} \pi_{2m+1} - \pi_{2m+1}
\partial_{mu} \pi_{2m}\right)^2 ,
\label{B1}
\end{eqnarray}
\narrowtext
where the indices $l$ and $m$  run from 1 to $2N-2$ and from 1 to $N-1$. The
values
of the couplings are the same as in (\ref{C1'}).
 As in the body of the paper, we focus on the situation
near the zero-temperature phase transition, which generally occurs at some
$g  = g_c (\gamma_{\mu})$. We also assume that the anisotropy
is small, and make all computations to leading order in $\gamma$ only.

The large $N$ expansion proceeds along the same lines as for
square-lattice antiferromagnets
{}~\cite{CSY}. We introduce the condensate value, $\sigma$, via
\begin{equation}
\sigma_0 = \sqrt{N} \sigma + \tilde{\sigma}_{0}
\label{B00}
\end{equation}
 and impose the constraint by introducing  the
Lagrange multiplier $\lambda$ into the functional integral.
At $N=\infty$, the saddle-point equation  gives
\begin{equation}
\sigma^2 = \frac{g_c - g}{g_c},
\label{B1A}
\end{equation}
 where $g_c$ is the same as in the isotropic case.

 We first consider
how the ratio of the stiffnesses scales with $g - g_c$. For this we observe
that
the only  anisotropic piece in (\ref{B1}) which survives
 at infinite $N$ is
$(\gamma_{\mu}/2g_{\mu})~
\sigma^2 ~(\partial_{\mu} \sigma_1)^2$. Hence, at $N= \infty$,
the propagators for $\pi$
fields remain the same as in the isotropic $O(2N)$ model
\begin{equation}
G_{\pi} = \frac{g_{\tau}}{2}~ \frac{1}{\omega^{2}_{n} + c^{2}_0 k^2},
\label{B2}
\end{equation}
while the propagator for the $\sigma_1$ field acquires a correction linear in
$\gamma_{\mu}$
\begin{equation}
G_{\sigma_1} = \frac{g_{\tau}}{2}~ \frac{1}{(1 + \gamma_{\tau} \sigma^2)
\omega^{2}_{n} + (1 + \gamma_x \sigma^2) c^{2}_0 k^2}
\label{B3}
\end{equation}
where $c_0 =
\sqrt{\rho^{0}_{\perp} /\chi^{0}_{\perp}}$.
Eqns (\ref{B2}) and (\ref{B3}) identify (upto an overall
factor) $G_{\pi}$ and $G_{\sigma_1}$ as the
transverse and longitudinal  propagators of gapless spin-wave excitations in
the ordered state. Each of the propagators can now be
reexpressed in terms of the
fully renormalized spin-stiffness  and  spin
susceptibility by collecting $\gamma$- independent terms, which are the same
for both propagators. Comparing then the
two expressions, we obtain
\begin{equation}
\frac{\rho_{\parallel} -
\rho_{\perp}}{\rho_{\perp}} =  \gamma_{x} \sigma^2 ~~~~~~~
 \frac{\chi_{\parallel} -
\chi_{\perp}}{\chi_{\perp}} =  \gamma_{\tau} \sigma^2
\label{B4}
\end{equation}
where $\rho$ and $\chi$ are now fully renormalized $T=0$ spin-stiffness and
spin susceptibility respectively.
We see from (\ref{B4}) that while the relative difference of
the bare stiffnesses
is $\gamma_{\mu}$, the  ratio of the renormalized
stiffnesses contains the extra factor $\sigma^2$ and therefore tends
to zero as the system approaches quantum phase transition point. We
now introduce the crossover exponents $\phi_1$ and $\phi_2$
in the same way as in Sec \ref{eff_act}. Namely, we
decompose $\gamma_{\mu}$ into their
trace and traceless parts as
\begin{equation}
 \gamma_x = \gamma_1 + \gamma_2 ~~~~ \gamma_{\tau} = \gamma_1 - 2 \gamma_2
\label{B5}
\end{equation}
 and define $\phi_1$ and $\phi_2$ as
\begin{eqnarray}
\frac{\rho_{\parallel} -
\rho_{\perp}}{\rho_{\perp}} &=&  \gamma_{1} (\xi_{J})^{-\phi_1} + \gamma_2
(\xi_J)^{-\phi_2} \nonumber\\
\frac{\chi_{\parallel} -
\chi_{\perp}}{\chi_{\perp}} &=&  \gamma_{1} (\xi_{J})^{-\phi_1} - 2\gamma_2
(\xi_J)^{-\phi_2}
\label{B6}
\end{eqnarray}
where $\xi_J$ is the Josephson correlation length measured in lattice units.
As $g$ approaches $g_c$, $\xi_J$ behaves as $\xi_J \sim (1 - g_x/g_c)^{-\nu}$.
At $N=\infty$, $\nu =1$. Using (\ref{B1A}), we then
 obtain $\phi_1 = \phi_2
=1$.

Our next step will be to calculate the  $1/N$ corrections to the crossover
exponents. The corresponding diagrams are presented in
Fig~\ref{figt=0cross}.
It is not difficult to show that the polarization operator at $T=0$,
$\Pi^{*}({k},i\omega)$, has no $\gamma$-dependent
corrections to the leading order in $1/N$ and we therefore can use
the same expression as in the isotropic case~\cite{CSY}:
\begin{equation}
\Pi^{*}({k},i\omega) = \frac{\hbar c^{2}_0}{8 \sqrt{c^{2}_0 k^2 + \omega^2}} +
\frac{2}{g_{\tau}} \, \langle\sigma\rangle^2 \,
\frac{1}{c^{2}_0 k^2 + \omega^2}
\label{B7}
\end{equation}
We will also need the expression for the propagator for the fluctuating
component of the field along the direction of the condensate:
\begin{equation}
G_{\tilde{\sigma}_0} = \frac{g_{\tau}}{2}~
\frac{1}{\omega^{2}_{n} + c^{2}_0 k^2},
\left(1 - \frac{2}{g} \, \langle\sigma\rangle^2 \,
\frac{1}{\Pi^{*}({k},i\omega)}~\frac{1}{\omega^{2}_{n} + c^{2}_0 k^2}
 \right),
\label{B8}
\end{equation}
and expressions for $\langle\sigma\rangle^2$ and $\xi_J$
with logarithmic accuracy to order $1/N$:
\begin{eqnarray}
\langle\sigma\rangle^2  &=& \left (\frac{g_{c} - g}{g_{c}} \right)\, \left(1 +
\frac{4}{\pi^2 N} \, \log \frac{g_{c}}{g_{c} - g} \right) \nonumber \\
\xi^{-1}_J &=& \left (\frac{g_{c} - g}{g_{c}} \right)\, \left(1 +
\frac{16}{3 \pi^2 N} \, \log \frac{g_{c}}{g_{c} - g} \right)  .
\label{B9}
\end{eqnarray}
The evaluation of the diagrams is now straightforward. Collecting the
contributions from all diagrams on Fig. \ref{figt=0cross}
and restricting to only logarithmic contributions to order $1/N$, we obtain
after some algebra
\begin{eqnarray}
\frac{\rho_{\parallel} -
\rho_{\perp}}{\rho_{\perp}} &=&  \frac{g_c -g}{g_c}~\left[\gamma_{x} \left(1
 + \frac{64 L}{15 \pi^2 N}\right) + \gamma_{\tau}~\frac{16 L}{15 \pi^2
N}\right]
 \nonumber\\
\frac{\chi_{\parallel} -
\chi_{\perp}}{\chi_{\perp}} &=&
\frac{g_c -g}{g_c}~\left[\gamma_{\tau} \left(1
 + \frac{16 L}{5 \pi^2 N}\right) + \gamma_{x}~\frac{32 L}{15 \pi^2 N}\right]
\label{B10}
\end{eqnarray}
where $L = \log(1 - g/g_c)$.
We then use (\ref{B5}) and find
\begin{equation}
\phi_1 = 1 + \frac{112}{15 \pi^2 N}~~~~~~\phi_2 = 1 + \frac{32}{3 \pi^2 N}
\label{B11}
\end{equation}
These values for the crossover exponents coincide
with Eqn (\ref{defphi1}) obtained by other means in Sec. \ref{eff_act}

Our next move will be to compute, in the $1/N$ expansion,
the critical exponent for the order parameter.
Using the definition (\ref{II6}) and Eqns (\ref{B0}) and (\ref{B00}), we obtain
\begin{equation}
N_0 =   S Z_S \left[\left(\langle \sigma \rangle^2 +
\frac{1}{N} \langle \tilde{\sigma}_{0} \rangle^2 \right) +
{\cal O}(\frac{1}{N})\right]
\label{B13}
\end{equation}
where
${\cal O}(1/N)$ stands for {\em regular} $1/N$ corrections from the
other components of $z$-field which can be neglected in the calculations of the
critical exponents. Using then (\ref{B8}) we find
\begin{equation}
N_0 =   S Z_S \langle \sigma \rangle^2
\left ( 1 - \frac{2}{g}~\int~\frac{1}{\Pi^{*}({k},i\omega)}~
\frac{1}{\omega^{2}_{n} + c^{2}_0 k^2} \right).
\label{B14}
\end{equation}
Performing the integration with the logarithmic accuracy and
exponentiating the result, we obtain
\begin{equation}
N_0 =
S Z_S  \langle \sigma \rangle^{\varepsilon}
\label{B15}
\end{equation}
where
\begin{equation}
\varepsilon = 1 +  \frac{4}{N \pi^2} .
\label{B16}
\end{equation}
Expressing now $\langle \sigma \rangle$ in a conventional way as
$\langle \sigma \rangle^2 = \left(\Delta g /g_c \right)^{\beta}$, where
$\beta = 1 - 4/N \pi^2$, we find $ N_0 \sim (\Delta
g/g_c)^{\bar{\beta}}$, where
\begin{equation}
{\bar \beta} = 1 + O\left(\frac{1}{N^2}\right)
\label{B16'}
\end{equation}

We will also need the $1/N$ result  for the staggered spin susceptibility
at $T=0$. Using (\ref{II5}), (\ref{B0}), (\ref{B00}) and the result obtained in
 Section \ref{ren_cl} that the number of transverse spin-wave modes in the
ordered state is $N_{sw} = 2 N$, we find that the transverse spin
susceptibility
at $N= \infty$ and  $k \ll \xi_{J}^{-1}$ is related to the propagator of the
$z$-field
\begin{equation}
\chi^{b b}_s (k, i\omega) =
\frac{(S Z_s)^2 \langle \sigma \rangle^2}{2 \rho^{0}_{\perp}}
{}~\frac{1}{k^2 + \omega^2 /c^{2}_0}.
\label{B18}
\end{equation}
where index $b$ labels transverse spin components. The relevant $1/N$
corrections are now the same as in collinear antiferromagnets.
Using the results of Ref~\cite{CSY}, we find that they can be completely
absorbed
into the renormalization of $N_0$ and $\rho_{\perp}$.
Assuming, as in the bulk of the paper,  that the anisotropic term in the action
will transform $\rho_{\perp}$ and $c_{\perp}$ into $\rho_s$ and $c$ given by
(\ref{C29}), we obtain
\begin{equation}
\chi^{b b}_s (k, i\omega) =
\frac{N^{2}_0}{2 \rho_{s}}
{}~\frac{1}{k^2 + \omega^2 /c^{2}}.
\label{B17}
\end{equation}

Finally, we calculate in $1/N$ expansion, the critical exponent $\bar \eta$.
For
this, we consider the behavior of dynamical spin susceptibility at $T=0$ right
at the transition point, $g =g_c$. At this point $\langle \sigma \rangle =0$,
and the spin-spin correlation function is again related to the polarization
operator of $z-$fields. At $N= \infty$ and $g = g_c$, the polarization operator
is given by the first term in (\ref{B7}). To first order in $1/N$, we have to
consider both self-energy and vertex corrections in the polarization bubble
(Fig \ref{figcorrpi}). They both
 contain logarithms of external momentum. Evaluating
the diagrams in Fig \ref{figcorrpi} with logarithmical accuracy, we obtain:
\begin{equation}
\Pi^{*}({k},i\omega) = \frac{A_{N}}{(c\Lambda)^{2 \mu}}~
\frac{c^2}{8 (c^2 k^2 + \omega^2)^{1 - {\bar \eta}/2}}
\label{B19}
\end{equation}
where $A_{N} = 1 + O(1/N)$, and
\begin{equation}
{\bar \eta} = 1 + \frac{32}{3 \pi^2 N}.
\label{B20}
\end{equation}

\begin{figure}
\caption{a. Self-energy corrections to spinon propagator.
The heavy solid line is a full spinon propagator, and the
wavy line is an inverse polarization operator for spinons.
The analytical expression for the vertex is given by (\protect\ref{C10}).
 b. Diagrams which contribute to current-current correlation
functions $<K^{a}_{\mu} K^{b}_{\mu}>$ and $<J_{\mu} J_{\mu}>$
to first order in $1/N$ {\em and} to first order in
$\gamma_{\mu}$. The side vertices (shaded) in the bubbles are $2 T^{ab}
k_{\mu} (1 + \gamma_{\mu})/g_{\mu}$ for $U(1)$ response and
$2 \delta^{ab} k_{\mu}(1 + \gamma_{\mu}/N)/g_{\mu}$ for $SU(N)$ response.}
\label{figcurrent}
\end{figure}

\begin{figure}
\caption{Diagrams which contribute to renormalization of the $\Gamma$ vertex
(\protect\ref{C10}) to order $1/N$.}
\label{figvertren}
\end{figure}

\begin{figure}
\caption{Diagrams which contribute to the renormalization of the
polarization operator to first order in $1/N$ at $\gamma_{\mu} =0$. As in
Fig.\protect\ref{figcurrent}, heavy solid lines are full spinon
propagators.}
\label{figcorrpi}
\end{figure}

\begin{figure}
\caption{Second-order self-energy corrections to magnon
propagators due to cubic vertices. Notice that cubic terms
always produce anomalous self-energy terms.}
\label{sec_or_dia}
\end{figure}

\begin{figure}
\caption{Self-energy diagrams to order $1/N$ for the propagator of
$\sigma_1$ field  in the ordered state at $T=0$. Solid line
is the propagator of $\sigma_1$ given by (\protect\ref{B3}),
dashed line represents the condensate $\sigma$,
wavy line is inverse polarization operator, and
heavy solid line is the propagator of $\tilde{\sigma}_0$ field
introduced in (\protect\ref{B00}): $G_{\tilde{\sigma}_0} = G_{\pi} (1 -
(2/g_{\tau}) \sigma^2 G_{\pi}/\Pi^{*})$, where $G_{\pi}$ is given by
(\protect\ref{B2}).}
\label{figt=0cross}
\end{figure}


\begin{references}
\bibitem{comment}  Numerical proof that the gap
at  at $k =\pm Q$ is absent was  first given in \cite{Rast_Tassi}.
Analytically, this was first shown
in \cite{Chub} in a way different from the one presented here.
\end{references}
\end{document}